 \documentclass[final,3p,11pt,authoryear]{elsarticle}

\makeatletter
\def\ps@pprintTitle{%
 \let\@oddhead\@empty
 \let\@evenhead\@empty
 \def\@oddfoot{\centerline{\thepage}}%
 \let\@evenfoot\@oddfoot}
\makeatother
\usepackage{color}
\usepackage[usenames,dvipsnames]{xcolor}
\usepackage{hyperref}

\usepackage{natbib}
\usepackage{graphicx}
\usepackage{caption}
\usepackage{subcaption}
\usepackage{amsfonts,amssymb,amsbsy,amsmath,mathtools}

\journal{International Journal of Solids and Structures}
\begin{document}

\begin{frontmatter}

\renewcommand{\thefootnote}{\fnsymbol{footnote}}
\title{\textbf{Two-scale micropolar plate model for web-core sandwich panels\let\thefootnote\relax\footnote{{\color{Blue}\textbf{Recompiled, unedited accepted manuscript}}. \copyright 2019. Made available under \href{https://creativecommons.org/licenses/by-nc-nd/4.0/}{{\color{Blue}\textbf{\underline{CC-BY-NC-ND 4.0}}}}}}}




\author[add1,add2]{Anssi T. Karttunen\corref{cor1}}
\cortext[cor1]{Corresponding author. anssi.karttunen@iki.fi. \textbf{Cite as}: \textit{Int. J. Solids Struct.} 2019;170:82--94 \href{https://doi.org/10.1016/j.ijsolstr.2019.04.026}{{\color{OliveGreen}\textbf{\underline{doi link}}}}}
\author[add2]{J.N. Reddy}
\address[add1]{Aalto University, Department of Mechanical Engineering, FI-00076 Aalto, Finland}
\address[add2]{Texas A\&M University, Department of Mechanical Engineering, College Station, TX 77843-3123, USA}
\author[add1]{Jani Romanoff}

\begin{abstract}
A 2-D micropolar equivalent single-layer (ESL), first-order shear deformation (FSDT) plate model for 3-D web-core sandwich panels is developed. First, a 3-D web-core unit cell is modeled by classical shell finite elements. A discrete-to-continuum transformation is applied to the microscale unit cell and its strain and kinetic energy densities are expressed in terms of the macroscale 2-D plate kinematics. The hyperelastic constitutive relations and the equations of motion (via Hamilton's principle) for the plate are derived by assuming energy equivalence between the 3-D unit cell and the 2-D plate. The Navier solution is developed for the 2-D micropolar ESL-FSDT plate model to study the bending, buckling, and free vibration of simply-supported web-core sandwich panels. In a line load bending problem, a 2-D classical ESL-FSDT plate model yields displacement errors of 34--175\% for face sheet thicknesses of 2--10 mm compared to a 3-D FE solution, whereas the 2-D micropolar model gives only small errors of 2.7--3.4\% as it can emulate the 3-D deformations better through non-classical antisymmetric shear behavior and local bending and twisting.
\end{abstract}

\end{frontmatter}


\section{Introduction}
Web-core sandwich panels consist of two horizontal face sheets separated by straight web plates that are aligned orthogonally to the faces. The core may be additionally filled with a low-density material such as a polymeric foam or balsa wood. The face sheets are responsible for taking bending and in-plane loads while the lightweight core carries the greater part of transverse shear loads. The primary function of any sandwich panel is usually to provide significant weight savings in comparison to more conventional structural configurations, such as stiffened solid plates, without compromising on the overall mechanical performance.

Two different web-core sandwich constructions are regularly encountered in the literature. Foam and wood-filled web-core panels with glass fiber reinforced polymer (GFRP) faces and webs have been designed for use in bridge decks and building roofs and floors \citep{zi2008,keller2008,huo2015,wang2015,zhu2018}. Laser-welded web-core steel sandwich panels, on the other hand, have been developed and used mainly in shipbuilding as staircase landings and non-structural walls \citep{roland1997,kujala2005}. The laser welds penetrate through the faces into the web-core and the panels may be filled with polymeric foams as well \citep{karttunen2017b}. Better understanding of their limit state behavior is making way for the steel panels to more challenging applications like ship decks, see, e.g., the works of \cite{kolsters2010,jelovica2013,jiang2014,frank2015} and \cite{korgesaar2016,korgesaar2018}. \cite{kujala2005} have estimated that ship decks constructed of steel sandwich panels offer 30--50\% weight savings compared to traditional stiffened steel plate configurations. This study is primarily motivated by all-steel web-core sandwich panels and ship structures. Nevertheless, such panels also show good potential for applications in bridges and buildings as evidenced by \cite{bright2004,bright2007,nilsson2017} and \cite{briscoe2011}. In addition, the methods presented in this paper should facilitate the structural analysis of GFRP sandwich panels in the future as well.

In order to analyze the global structural response of a large cruise ship or some other megastructure within computational limits, the main structural components need to be modeled in a homogenized sense without accounting for every small detail. To this end, a sandwich panel used for a ship or bridge deck may be modeled as an equivalent single-layer (ESL) beam or plate based on the first-order shear deformation theory (FSDT) \citep{reddy2004}. Lately, different ESL-FSDT models have been developed for web-core sandwich beams using classical \citep{romanoff2007b}, couple-stress \citep{romanoff2016,gesualdo2017,penta2017}, and micropolar \citep{karttunen2018a,karttunen2019a} continuum theories. Recently, analytical solutions founded on discrete classical models have also been formulated for web-core plates \citep{pydah2016,pydah2017,pydah2018}. As for the different ESL-FSDT beam models, it has been shown that only the micropolar approach can capture accurately the deformations of a web-core beam because it considers both symmetric and antisymmetric shear and local bending deformations \citep{karttunen2018a,karttunen2019a}. This is explained schematically in Fig.~1 in the case of a 3-D web-core plate. In this paper, we develop a novel 2-D micropolar ESL-FSDT plate theory for web-core sandwich panels such as the one shown in Fig.~1.
\begin{figure}[hb]
\centering
\includegraphics[trim={0mm, 3mm, 0mm, 3mm},scale=0.69]{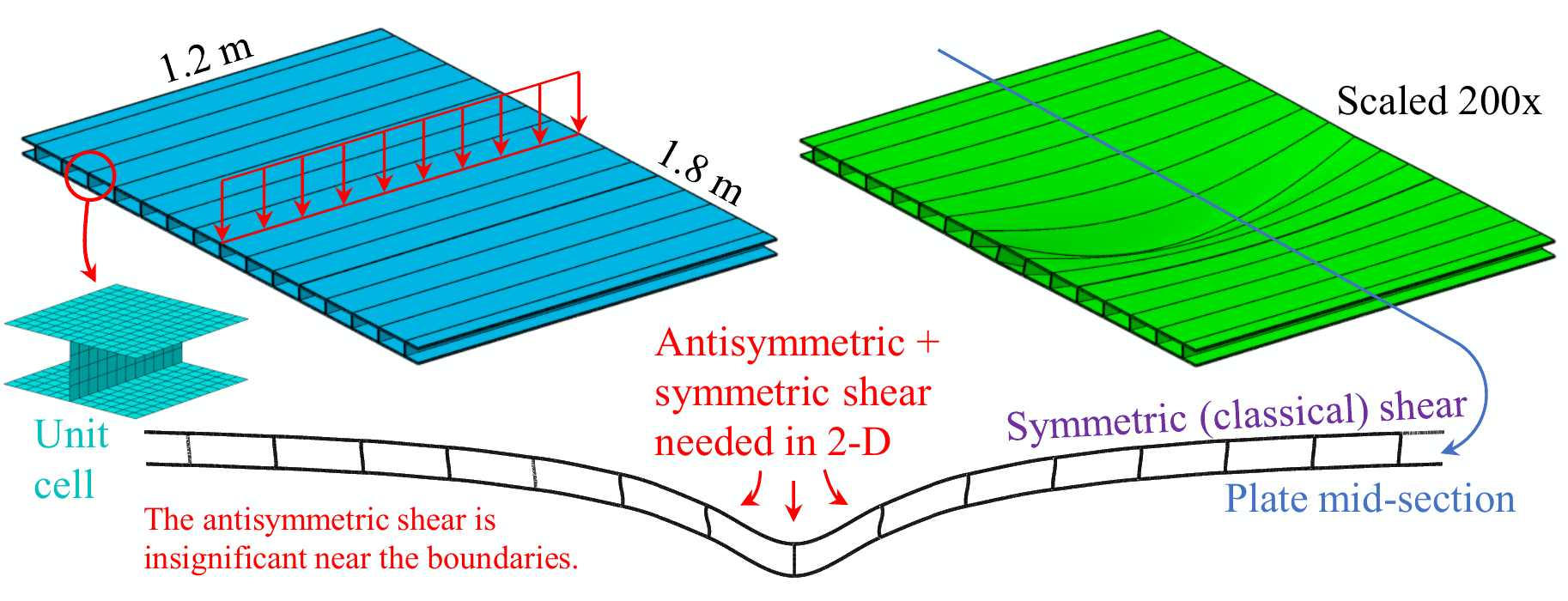}
\caption{Simply-supported all-steel web-core sandwich panel under a line load modeled in 3-D by shell finite elements (Section 4). The 2-D micropolar plate theory to be developed allows antisymmetric shear deformation to emerge at locations where the 3-D deformations cannot be reduced to 2-D by considering only symmetric shear behavior. Local bending can also be seen within the unit cells near the plate center but not in the vicinity of the supported edges.}
\end{figure}

In the micropolar continuum theory, each material particle has, in addition to the three classical displacements, an orientation described by three rotational degrees of freedom  \citep{eringen2012}. Thus, with six independent degrees of freedom, every material point of a micropolar continuum is phenomenologically equivalent to a rigid body. The micropolar theory is a simplification of the micromorphic theory where each point has three deformable directors and includes nine additional degrees of freedom compared to the classical theory \citep{eringen2012}. The works of \cite{eringen2012}, \cite{altenbach2009} and \cite{altenbach2010} discuss the key theoretical aspects of micropolar plate theories and the literature reviews therein encompass a lengthy period of research on such plates. Of the more recent papers on micropolar plates we mention those of \cite{kvasov2013,sargsyan2014,folkow2015,ansari2017} and \cite{zozulya2018}. However, the aforementioned models are not applicable in the present context as their constitutive equations do not address any particular microstructure such as a web-core. As pointed out by \cite{altenbach2010}: \textit{The main problem of any micropolar theory is the establishment of the constitutive equations}. In this paper, our focus is on developing a two-scale micropolar constitutive modeling aprroach for web-core sandwich plates. The method is similar to an approach that has been used for beam-like lattices and plate-like grids, but not for sandwich plates \citep{noor1980,noor1988,ostoja2002,karttunen2019a}.

In a nutshell, the present two-scale constitutive modeling approach for a 2-D micropolar ESL-FSDT plate is based on ideas akin to those behind second-order computational homogenization techniques \citep{kouznetsova2002,larsson2007}: (1) No constitutive model is assumed for the 2-D \textit{macroscale} ESL-FSDT plate a priori. (2) Instead, an I-shaped, \textit{microscale} web-core unit cell is modeled by classical constituents (i.e., by shell finite elements, see Fig.~1). (3) Then the polar macroscale plate kinematics are imposed on the microscale unit cell in order to bridge the two scales. However, instead of solving a nested boundary value problem as in computational homogenization, (4) the (hyperelastic) constitutive relations for the 2-D plate will be determined directly from the unit cell strain energy which at this point is given in terms of the macroscale ESL-FSDT plate strains. In addition to the strain energy, the kinetic energy of the unit cell is obtained through a similar procedure after which the equations of motion for the 2-D micropolar ESL-FSDT web-core plate are derived via Hamilton's principle. Thus, the whole formulation of the micropolar plate theory is actually built upon on the constitutive modeling.

The fundamental reason for using the micropolar theory in the above modeling procedure is that it allows us to pass information on both displacements \textit{and} rotations from the discrete unit cell into an ESL plate continuum. After all, the classical shell finite elements used to model the discrete unit cell have both translational and rotational degrees of freedom. This way we are able to account, in addition to the antisymmetric shear deformations, for the local twisting and bending of the sandwich face sheets and webs with respect to their own mid-surfaces through couple-stress moments which is not possible in a conventional ESL-FSDT sandwich plate theory. The local face bending is well displayed near the center of the plate mid-section in Fig.~1 and it will be seen later that this local behavior is coupled through equilibrium to the antisymmetric shear behavior. In a broad sense, the two-scale energy approach for the 2-D micropolar plate is another step towards a general constitutive modeling technique in micropolar elasticity with particular emphasis on mid-surface structural components such as beams, plates and shells made of lattice materials.

The rest of the paper is organized as follows. The two-scale constitutive modeling for a web-core sandwich panel which gives the stress resultant equations for the 2-D micropolar plate is carried out in Section 2. Along with the constitutive equations, the procedure yields the strain and kinetic energy densities of a unit cell which are used in Section 3 to derive the governing equations of the novel 2-D micropolar ESL-FSDT plate theory in terms of the stress resultants by employing Hamilton's principle. The Navier solution is developed for the plate in order to study numerically static bending, linear buckling and free vibration problems of simply-supported web-core sandwich panels in Section 4. Concluding remarks are given in Section 5.
\section{Two-scale constitutive modeling}
\subsection{Micropolar displacements and strains}
The 3-D displacements and microrotations of a plate-like continuum can be approximated by 2-D mid-surface kinematic variables ($u_x, u_y, u_z, \phi_x, \phi_y, \psi_x, \psi_y$) so that
\begin{equation}
\begin{aligned}
U_x(x,y,z,t)&=u_x(x,y,t)+z\phi_x(x,y,t), \\
U_y(x,y,z,t)&=u_y(x,y,t)+z\phi_y(x,y,t), \\
U_z(x,y,z,t)&=u_z(x,y,t), \\
\Psi_x(x,y,z,t)&=\psi_x(x,y,t), \\
\Psi_y(x,y,z,t)&=\psi_y(x,y,t), \\
\Psi_z(x,y,z,t)&=0,
\end{aligned}
\end{equation}
where $t$ is time, $(u_x, u_y, u_z)$ denote the displacements of a point on the plane $z=0$, and ($\phi_x,\phi_y$) are the rotations of a transverse normal about the $y$- and $x$-axes, respectively, whereas ($\psi_x,\psi_y$) are microrotations about the $x$- and $y$-axes, respectively. Finally, note the following two assumptions that are introduced here for the micropolar plate theory: (1) the formulation is for plates of constant thickness, which (2) do not possess a drilling degree of freedom [$\Psi_z(x,y,z)=0$].

Following the theory of micropolar elasticity \citep{eringen2012}, the nonzero strains of the micropolar plate are
\begin{equation}
\begin{aligned}
\epsilon_{xx}&=U_{x,x}=u_{x,x}+z\phi_{x,x}=\epsilon_{xx}^0+z\kappa_{xx}, \\
\epsilon_{yy}&=U_{y,y}=u_{y,y}+z\phi_{y,y}=\epsilon_{yy}^0+z\kappa_{yy}, \\
\epsilon_{xy}&=U_{y,x}-\Psi_z=u_{y,x}+z\phi_{y,x}=\epsilon_{xy}^0+z\kappa_{xy}, \\ \epsilon_{yx}&=U_{x,y}+\Psi_z=u_{x,y}+z\phi_{x,y}=\epsilon_{yx}^0+z\kappa_{yx}, \\
\epsilon_{xz}&=U_{z,x}+\Psi_y=u_{z,x}+\psi_{y}, \\
\epsilon_{zx}&=U_{x,z}-\Psi_y=\phi_{x}-\psi_{y}, \\
\epsilon_{yz}&=U_{z,y}-\Psi_x=u_{z,y}-\psi_{x}, \\
\epsilon_{zy}&=U_{y,z}+\Psi_x=\phi_{y}+\psi_{x}, \\
\chi_{xx}&=\Psi_{x,x}=\psi_{x,x}, \, \chi_{yy}=\Psi_{y,y}=\psi_{y,y},  \\
\chi_{xy}&=\Psi_{y,x}=\psi_{y,x}, \, \chi_{yx}=\Psi_{x,y}=\psi_{x,y},
\end{aligned}
\end{equation}
where $x$ and $y$ in the subscripts after the comma denote partial differentiation with respect to coordinates $x$ and $y$, respectively. The symmetric shear strains are defined as 
\begin{equation}
\begin{aligned}
\gamma^s_x&=\epsilon_{xz}+\epsilon_{zx}=u_{z,x}+\phi_x, \\
\gamma^s_y&=\epsilon_{yz}+\epsilon_{zy}=u_{z,y}+\phi_y
\end{aligned}
\end{equation}
and the antisymmetric shear strains are
\begin{equation}
\begin{aligned}
\gamma^a_x&=\epsilon_{xz}-\epsilon_{zx}=u_{z,x}-\phi_x+2\psi_y=2(\psi_y-\omega_{y}), \\
 \gamma^a_y&=\epsilon_{yz}-\epsilon_{zy}=u_{z,y}-\phi_y-2\psi_x=2(\omega_{x}-\psi_x),
\end{aligned}
\end{equation}
where $(\omega_{x},\omega_{y})$ are the macrorotations. The symmetric shear strains $(\gamma^s_x,\gamma^s_y)$ take the same forms as the shear strains in the conventional ESL-FSDT (ESL-Mindlin) plate theory founded on classical elasticity. The antisymmetric parts are defined by the macrorotations and the microrotations. In addition to having  rotational degrees of freedom independent of the translational displacements, the micropolar plate can transmit couple-stress as well as the usual, classical force-stresses. 
\subsection{Discrete-to-continuum transformation}
\begin{figure}
\centering
\includegraphics[scale=0.55]{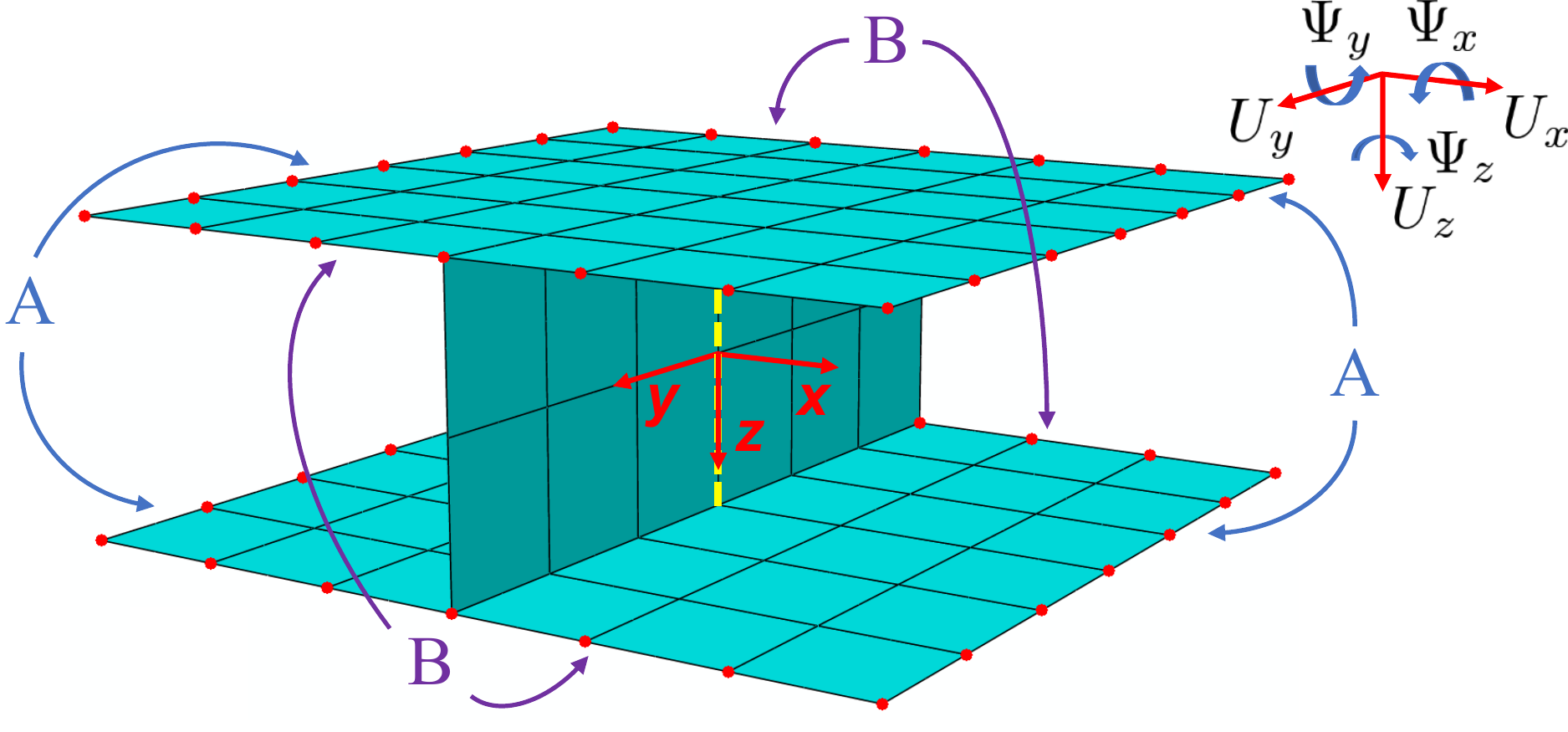}
\caption{Web-core unit cell modeled by shell finite elements in Abaqus. The mesh is coarse for illustrative purposes. In essence, the 3-D unit cell represents a lattice material and is analogous to an infinitesimal material element. In generating a statically condensed FE model for the two-scale constitutive modeling approach, certain DOFs are retained on face edges A and B while all interior DOFs are condensed out. No boundary conditions are applied.}
\end{figure}
Figure 2 shows, for modeling purposes, a web-core unit cell attached to an arbitrary \textit{fiber} of a micropolar plate; see the dashed line along $z$-axis for the fiber. This situation is analogous to a case in which a unit cell is attached to an arbitrary cross section of a micropolar beam \citep{karttunen2019a}. The 2-D micropolar continuum plate as a whole is a \textit{macrostructure} and the discrete 3-D unit cell represents its periodic \textit{microstructure}. In order to obtain the constitutive equations for the plate, the strain energy of the discrete microscale FE unit cell needs to be expressed in terms of the continuous macroscale strains (2)--(4) so as to bridge the two scales.

Static condensation is applied at all nodes of the FE unit cell in Fig.~2 to the global drilling degree of freedom, i.e., to the rotation with respect to $z$-axis. Certain nodal DOFs are retained only at the highlighted face edge nodes of the unit cell as will be discussed in detail in Section 2.3. The practical details of the following modeling procedures that relate to the use of Abaqus and Matlab are discussed in Appendix A. But first, in general, the strain energy of a web-core unit cell after static condensation is
\begin{equation}
W=\frac{1}{2}\mathbf{d}^{\textrm{T}}\mathbf{k}\mathbf{d},
\end{equation}
where $\mathbf{d}$ contains the retained degrees of freedom and $\mathbf{k}$ is the corresponding unit cell stiffness matrix. Next the discrete nodal displacements at the unit cell edges are expressed in terms of the plate (fiber) displacements (1). With distance from an arbitrary fiber located within a micropolar plate domain, Taylor series expansion of Eqs.~(1) for edge nodes $i=1,2,\ldots,n$ on the bottom and top faces $(z=\pm h/2)$ leads to
\begin{equation}
\begin{aligned}
U_x^i&=u_x\pm\frac{h}{2}\left[\frac{1}{2}\left(\gamma^s_x-\gamma^a_x\right)+\psi_y\right]+\Delta x^i\left[\epsilon^0_{xx}\pm\frac{h}{2}\kappa_{xx}\right]+\Delta y^i\left[\epsilon^0_{yx}\pm\frac{h}{2}\kappa_{yx}\right], \\
U_y^i&=u_y\pm\frac{h}{2}\left[\frac{1}{2}\left(\gamma^s_y-\gamma^a_y\right)-\psi_x\right]+\Delta y^i\left[\epsilon^0_{yy}\pm\frac{h}{2}\kappa_{yy}\right]+\Delta x^i\left[\epsilon^0_{xy}\pm\frac{h}{2}\kappa_{xy}\right], \\
U_z^i&=u_z+\Delta x^i\left[\frac{1}{2}\left(\gamma^s_x+\gamma^a_x\right)-\psi_y\right]+\Delta y^i\left[\frac{1}{2}\left(\gamma^s_y+\gamma^a_y\right)+\psi_x\right], \\
\Psi^i_x&=\psi_x+\Delta x^i\chi_{xx}+\Delta y^i\chi_{yx}, \\
\Psi^i_y&=\psi_y+\Delta x^i\chi_{xy}+\Delta y^i\chi_{yy},
\end{aligned}
\end{equation}
where $\Delta x^i$ and $\Delta y^i$ are the nodal coordinates of the FE unit cell in the present context. Moreover, the micropolar plate strains (2)--(4) have been imposed on the rotations $(\phi_x,\phi_y)$ and the gradient terms. The expansions (6)$_4$ and (6)$_5$ of the rotational degrees of freedom of the 3-D FE unit cell modeled by shell elements are written completely in terms  micropolar features that do not appear in classical continuum theories, that is, in terms of the microrotations $\psi_x$ and $\psi_y$ and their derivatives. Recall also that $\Psi^i_z$ was condensed out. We can write the \textit{discrete-to-continuum transformation} (6) for nodes $i=1,2,\ldots,n$ in matrix form
\begin{equation}
\mathbf{d}=\mathbf{T}^{\phantom{ }}_u\mathbf{u}+\mathbf{T}^{\phantom{ }}_\epsilon\boldsymbol{\epsilon},
\end{equation}
where the vector for the continuous displacements is
\begin{equation}
\mathbf{u}=\left\{u_x \ \ u_y \ \ u_z \ \ \phi_x \ \ \psi_y \ \ \phi_y \ \ \psi_x \right\}^{\textrm{T}}.
\end{equation}
For the strains we have
\begin{equation}
\boldsymbol{\epsilon}=
\begin{Bmatrix}
\boldsymbol{\epsilon}^0 \\
\boldsymbol{\kappa} \\
\boldsymbol{\gamma} \\
\boldsymbol{\chi}
\end{Bmatrix},
\end{equation}
where
\begin{equation}
\boldsymbol{\epsilon}^0=
\begin{Bmatrix}
\epsilon_{xx}^0 \\ \epsilon_{yy}^0 \\ \epsilon_{xy}^0 \\ \epsilon_{yx}^0
\end{Bmatrix}, \quad
\boldsymbol{\kappa}=
\begin{Bmatrix}
\kappa_{xx} \\ \kappa_{yy} \\ \kappa_{xy} \\ \kappa_{yx}
\end{Bmatrix}, \quad
\boldsymbol{\gamma}=
\begin{Bmatrix}
\gamma^s_x \\ \gamma^a_x \\ \gamma^s_y \\ \gamma^a_y
\end{Bmatrix}, \quad
\boldsymbol{\chi}=
\begin{Bmatrix}
\chi_{xx} \\ \chi_{yy} \\ \chi_{xy} \\ \chi_{yx}
\end{Bmatrix},
\end{equation}
and $\mathbf{T}^{\phantom{ }}_u$ and $\mathbf{T}^{\phantom{ }}_\epsilon$ are transformation matrices. By applying the transformation (7) to the strain energy (5), it is straightforward to verify in all the cases studied in this paper that the displacement vector $\mathbf{u}$ (rigid body motion) does not contribute to the strain energy. We obtain
\begin{equation}
W=\frac{1}{2}\boldsymbol{\epsilon}^{\textrm{T}}\mathbf{T}_\epsilon^{\textrm{T}}\mathbf{k}\mathbf{T}^{\phantom{ }}_\epsilon\boldsymbol{\epsilon}.
\end{equation}
We define the \textit{areal density} of the unit cell strain energy as
\begin{equation}
W_0^A\equiv\frac{W}{A}=\frac{1}{2}\boldsymbol{\epsilon}^{\textrm{T}}\mathbf{C}\boldsymbol{\epsilon},
\end{equation}
where $A$ is the planform area of the unit cell. The constitutive matrix is given by
\begin{equation}
\mathbf{C}=\frac{1}{A}\mathbf{T}_\epsilon^{\textrm{T}}\mathbf{k}\mathbf{T}^{\phantom{ }}_\epsilon.
\end{equation}
The unit cell represents a \textit{lattice material} of which the micropolar plate is made of. Therefore, in analogy with any hyperelastic material, we write for the micropolar plate continuum
\begin{equation}
\mathbf{S}\equiv\frac{\partial W_0^A}{\partial \boldsymbol{\epsilon}}=\mathbf{C}\boldsymbol{\epsilon},
\end{equation}
where $\mathbf{S}$ is now the stress resultant vector of the micropolar plate. By employing Eq.~(14), the bridging of the two scales is founded on an assumption of strain energy equivalence between the macrostructure (plate) and the microstructure (unit cell). The explicit matrix form of Eq.~(14) is
\begin{equation}
\begin{Bmatrix}
\mathbf{N} \\
\mathbf{M} \\
\mathbf{Q} \\
\mathbf{P}
\end{Bmatrix}
=
\begin{bmatrix}
 \mathbf{A} & \mathbf{0} & \mathbf{0} & \mathbf{0}  \\
 \mathbf{0} & \mathbf{D} & \mathbf{0} & \mathbf{0}  \\
 \mathbf{0} & \mathbf{0} & \mathbf{G} & \mathbf{0}  \\
 \mathbf{0} & \mathbf{0} & \mathbf{0} & \mathbf{H}  \\
\end{bmatrix}
\begin{Bmatrix}
\boldsymbol{\epsilon}^0 \\
\boldsymbol{\kappa} \\
\boldsymbol{\gamma} \\
\boldsymbol{\chi}
\end{Bmatrix},
\end{equation}
where the vectors for the membrane $\mathbf{N}$, global bending and twisting $\mathbf{M}$, symmetric and antisymmetric shear $\mathbf{Q}$ and local (couple-stress related) bending and twisting $\mathbf{P}$ resultants read
\begin{equation}
\begin{aligned}
\mathbf{N}&=\left\{N_{xx} \ \ N_{yy} \ \ N_{xy} \ \ N_{yx} \right\}^{\textrm{T}}, \\
\mathbf{M}&=\left\{M_{xx} \ \ M_{yy} \ \ M_{xy} \ \ M_{yx} \right\}^{\textrm{T}}, \\
\mathbf{Q}&=\left\{Q^s_{x} \ \ Q^a_{x} \ \ Q^s_{y} \ \ Q^a_{y} \right\}^{\textrm{T}}, \\
\mathbf{P}&=\left\{P_{xx} \ \ P_{yy} \ \ P_{xy} \ \ P_{yx} \right\}^{\textrm{T}},
\end{aligned}
\end{equation}
respectively. The submatrices for the constitutive parameters in this study are
\begin{align}
&\mathbf{A}=
\begin{bmatrix}
 A_{11} & A_{12} & 0 & 0  \\
 A_{12} & A_{22} & 0 & 0  \\
 0 & 0 & A_{33} & A_{34}  \\
 0 & 0 & A_{34} & A_{44}  \\
\end{bmatrix},
\quad
\mathbf{D}=
\begin{bmatrix}
 D_{11} & D_{12} & 0 & 0  \\
 D_{12} & D_{22} & 0 & 0  \\
 0 & 0 & D_{33} & D_{34}  \\
 0 & 0 & D_{34} & D_{44}  \\
\end{bmatrix}, \\
&\mathbf{G}=
\begin{bmatrix}
 G_{11} & G_{12} & 0 & 0  \\
 G_{12} & G_{22} & 0 & 0  \\
 0 & 0 & G_{33} & G_{34}  \\
 0 & 0 & G_{34} & G_{44}  \\
\end{bmatrix},
\quad
\mathbf{H}=
\begin{bmatrix}
 H_{11} & H_{12} & 0 & 0  \\
 H_{12} & H_{22} & 0 & 0  \\
 0 & 0 & H_{33} & H_{34}  \\
 0 & 0 & H_{34} & H_{44}  \\
\end{bmatrix}.
\end{align}
The matrices include 24 constitutive parameters and are symmetric in all cases in this study. The constitutive modeling approach resulting in Eq.~(15) was presented in a quite general form and it is meant to be independent of the sandwich core; the discrete-to-continuum transformation (7) is applied  along the face edges after static condensation but not within the core. The approach is explored further in the next section.
\subsection{Constitutive parameters for a web-core panel}
The detailed constitutive modeling of the web-core unit cell is carried out in two separate phases which are outlined below. First, the global drilling degree of freedom is condensed out at every node of the FE model shown in Fig.~2, and the highlighted face edge nodes are taken as the master nodes. In addition, the following steps are taken.
\begin{itemize}
    \item[(1)] All the remaining five DOFs are retained at the master nodes while all DOFs are condensed out in the other (slave) nodes. The transformation (7) is applied to the strain energy of the resulting FE unit cell. However, the nodal rotations of the FE model are expanded only by
\begin{equation}
\Psi^i_x=\psi_x, \quad \Psi^i_y=\psi_y
\end{equation}
so that when Eq.~(14) is finally applied, we obtain $\mathbf{N}=\mathbf{A}\boldsymbol{\epsilon}^0, \mathbf{M}=\mathbf{D}\boldsymbol{\kappa}$ and  $\mathbf{Q}=\mathbf{G}\boldsymbol{\gamma}$ but $\mathbf{P}=\mathbf{H}\boldsymbol{\chi}$ for local bending and twisting remains to be derived. Expansions (19) are enough to fully account for the shear behavior, see Eq.~(4).
\item [(2)] As for $\mathbf{H}$, only the rotation $\Psi_y^i$ with respect to $y$-axis is retained on face edges A (including the corners), whereas on edges B only the rotation $\Psi_x^i$ with respect to $x$-axis is retained. The strain energy expression (11) uses the expansions
\begin{align}
\Psi^i_x&=\psi_x+\Delta x^i\chi_{xx}+\Delta y^i\chi_{yx}, \\
\Psi^i_y&=\psi_y+\Delta x^i\chi_{xy}+\Delta y^i\chi_{yy}.
\end{align}
In this case, Eq.~(14) gives the relation $\mathbf{P}=\mathbf{H}\boldsymbol{\chi}$.
\end{itemize}
\begin{figure}
\centering
\includegraphics[scale=1.2]{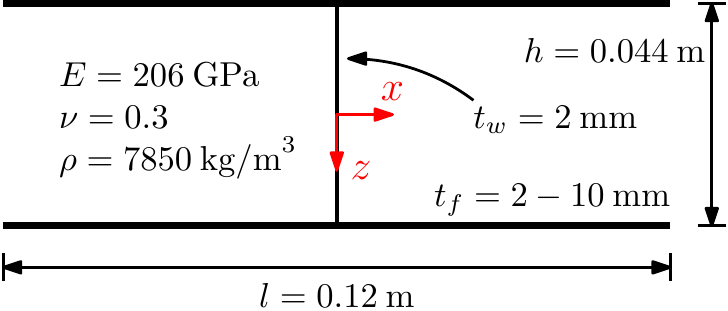}
\caption{Parameters of a web-core unit cell made of steel with Young's modulus $E=206$ GPa, Poisson ratio $\nu=0.3$ and density $\rho=7850$ kg/m$^3$. The length and width of the unit cell are both taken to be $l=0.12$ m so that the unit cell planform area is $A=l^2=0.0144$ m$^2$.}
\end{figure}

Figure 3 shows the web-core unit cell of Fig.~2 in the $x$-$z$-plane. The unit cell is made of steel and the face sheets are always of equal thickness. The constitutive parameters will be derived for face thicknesses $t_f=2,4,\ldots,10$ mm and for a web thickness of $t_w=2$ mm. The length and width of the unit cell are both $l=0.12$ m and the distance between the face sheet mid-surfaces is $h=0.044$ m. The face-web connections are rigid. 
The constitutive parameters derived from cases (1) and (2) are given in Table 1 for varying face sheet thickness $t_f$. In the case of the web-core panel it turns out that
\begin{equation}
A_{33}=A_{34}=A_{44} \quad \textrm{and} \quad D_{33}=D_{34}=D_{44}
\end{equation}
and, thus, only $A_{33}$ and $D_{33}$ are given in Table 1. It follows that 
\begin{equation}
N_{xy}=N_{yx} \quad \textrm{and} \quad M_{xy}=M_{yx}.
\end{equation}
The symmetry (23) will be utilized in deriving the equations of motion for the micropolar plate theory in Section 3. By constructing constitutive matrices $\mathbf{C}$ for different face thicknesses $t_f$ using the values given in Table 1, it is easy to verify that all eigenvalues of all constitutive matrices are positive. This means that the matrices are positive-definite. It follows that the web-core lattice material is stable in the conventional sense (i.e., strain energy $W$ is positive for nonzero strains $\boldsymbol{\epsilon}$).

Although only numerical values are obtained for the micropolar constitutive parameters, it is relatively easy to deduce by comparison the analytical forms of some of the parameters by the aid of the classical sandwich theory \citep{allen1969,vinson2018}. For the membrane part we have, to a very good approximation,
\begin{equation}
A_{11}\approx\frac{2Et_f}{1-\nu^2}, \quad A_{12}\approx\nu A_{11}, \quad A_{33}\approx\frac{Et_f}{1+\nu}\, ,
\end{equation}
which are all independent of the web thickness $t_w$. The value of $A_{22}$ is always slightly higher than that of $A_{11}$ because of the membrane stiffness provided by the web aligned with the $y$-direction. For the global bending and twisting stiffnesses we have
\begin{equation}
D_{11}\approx\frac{Et_fh^2}{2(1-\nu^2)}, \quad D_{12}\approx\nu D_{11}, \quad D_{33}\approx\frac{Et_fh^2}{4(1+\nu)}
\end{equation}
and analogously to $A_{22}$, $D_{22}$ is always a bit higher than $D_{11}$ because of the membrane stiffness provided by the web in the $y$-direction. As for the local twisting and bending, $H_{11}$, $H_{12}$ and $H_{22}$ are for twisting. For local bending we have
\begin{equation}
H_{33}\approx \frac{2EI_f}{(1-\nu^2)}=\frac{Et_f^3}{6(1-\nu^2)}\, ,
\end{equation}
which is the sum of the bending rigidities of the two faces with respect to their own mid-surfaces. From Table 1, we see that $H_{44}$ is considerably larger than $H_{33}$ because of the additional bending stiffness provided by the web in the $y$-direction. The coupling term $H_{34}$ for local bending is the only parameter that takes a negative value in the case of the web-core unit cell. For further details on the difference between global and local bending and twisting in sandwich structures see the book by \cite{allen1969} and a recent paper on web-core sandwich beams \citep{karttunen2018a}. The shear matrix $\mathbf{G}$ is not as straightforward to interpret. We only note that the 2-D micropolar ESL-FSDT plate theory does not employ any extrinsic shear correction factors. Equations (24)--(26) reflect the fact that all the constitutive parameters stem from the 3-D unit cell modeled by classical shell elements and the 2-D micropolar plate does not include any non-classical length scale parameters. The overall validity of the obtained constitutive parameters will be studied by numerical examples in Section 4.  The presented constitutive modeling approach also works in the context of classical elasticity, see Appendix B.
\begin{table}[ht]
\caption{Constitutive coefficients for web-core sandwich panels modeled as 2-D micropolar ESL-FSDT plates. See Fig.~3 for the unit cell properties.}
\begin{center}
\small
\begin{tabular}{{c}|{c}{c}{c}{c}{c}}
\hline
$\mathbf{A}$ [MN/m] & $t_f=2$ mm & $4$ mm & $6$ mm & $8$ mm & $10$ mm \\
\hline
$A_{11}$ & 905.495 & 1810.99 & 2716.48 & 3621.98 & 4527.47  \\
$A_{12}$ & 271.648 & 543.297 & 814.945 & 1086.59 & 1358.24  \\
$A_{22}$ & 997.277 & 1907.18 & 2814.44 & 3720.98 & 4627.24  \\
$A_{33}$ & 316.923 & 633.846 & 950.769 & 1267.69 & 1584.62  \\
\hline
$\mathbf{D}$ [MNm] & $t_f=2$ mm & $4$ mm & $6$ mm & $8$ mm & $10$ mm \\
\hline
$D_{11}$ & 0.43826 & 0.87652 & 1.31478 & 1.75304 & 2.19130  \\
$D_{12}$ & 0.13148 & 0.26296 & 0.39443 & 0.52591 & 0.65739  \\
$D_{22}$ & 0.46020 & 0.90585 & 1.35350 & 1.80011 & 2.24448  \\
$D_{33}$ & 0.15339 & 0.30678 & 0.46017 & 0.61356 & 0.76695  \\
\hline
$\mathbf{G}$ [MN/m] & $t_f=2$ mm & $4$ mm & $6$ mm & $8$ mm & $10$ mm \\
\hline
$G_{11}$ & 1.16557 & 4.26884 & 9.82709 & 18.1926 & 29.5600  \\
$G_{12}$ & 0.93837 & 3.92462 & 9.46343 & 17.8250 & 29.1924  \\
$G_{22}$ & 0.91865 & 3.89135 & 9.42900 & 17.7917 & 29.1609  \\
$G_{33}$ & 34.9111 & 45.7777 & 55.5037 & 66.3705 & 79.3196  \\
$G_{34}$ & 1.32756 & 5.01625 & 11.1895 & 20.0058 & 31.6599  \\
$G_{44}$ & 0.98325 & 4.17269 & 9.96295 & 18.5242 & 30.0194  \\
\hline
$\mathbf{H}$ [Nm] & $t_f=2$ mm & $4$ mm & $6$ mm & $8$ mm & $10$ mm \\
\hline
$H_{11}$ & 190.167 & 1416.73 & 4647.49 & 10763.7 & 20571.6  \\
$H_{12}$ & 23.8172 & 211.894 & 712.203 & 1656.92 & 3155.87  \\
$H_{22}$ & 181.960 & 1407.22 & 4638.13 & 10754.7 & 20563.1  \\
$H_{33}$ & 301.832 & 2414.65 & 8149.45 & 19317.2 & 37728.9  \\
$H_{34}$ &-90.5495 &-724.396 &-2444.84 &-5795.16 &-11318.7  \\
$H_{44}$ & 1317.40 & 7712.00 & 20232.4 & 38648.9 & 63268.3  \\
\hline
\end{tabular}
\end{center}
\end{table}
\subsection{Kinetic energy of web-core unit cell}
To obtain the kinetic energy expression for the 2-D micropolar ESL-FSDT web-core plate, only the velocity terms $(\dot{u}_x,\dot{u}_y,\dot{u}_z,\dot{\phi}_x,\dot{\psi}_y,\dot{\phi}_y,\dot{\psi}_x)$ are included in the expansions (6) [i.e., the gradients are not included and the strains are not imposed in Eq.~(6)]. The discrete-to-continuum transformation of the kinetic energy of a web-core unit cell can then be written as
\begin{equation}
\Tilde{K}=\frac{1}{2}\mathbf{\dot{d}}^{\textrm{T}}\mathbf{m}\mathbf{\dot{d}}=\frac{1}{2}\mathbf{\dot{u}}^{\textrm{T}}\mathbf{T}_{\dot{u}}^{\textrm{T}}\mathbf{m}\mathbf{T}^{\phantom{ }}_{\dot{u}}\mathbf{\dot{u}},
\end{equation}
where the dot on the variables denotes differentiation with respect to time and $\mathbf{T}^{\phantom{ }}_{\dot{u}}$ is a transformation matrix. The mass matrix $\mathbf{m}$ of the unit cell is obtained using shell finite elements (Appendix A) and by retaining the DOFs in the same manner as in modeling phase (1) which was explained in Section 2.3. We define the areal density of the unit cell kinetic energy as
\begin{equation}
\Tilde{K}_0^A\equiv\frac{\Tilde{K}}{A}=\frac{1}{2}\mathbf{\dot{u}}^{\textrm{T}}\mathbf{M}\mathbf{\dot{u}},
\end{equation}
where
\begin{equation}
\mathbf{M}=\frac{1}{A}\mathbf{T}_{\dot{u}}^{\textrm{T}}\mathbf{m}\mathbf{T}^{\phantom{ }}_{\dot{u}}=
\begin{bmatrix}
 m_{11} & 0 & 0 & 0 & 0 & 0 & 0 \\
 0 & m_{22} & 0 & 0 & 0 & 0 & 0 \\
 0 & 0 & m_{33} & 0 & 0 & 0 & 0 \\
 0 & 0 & 0 & m_{44} & m_{45} & 0 & 0 \\
 0 & 0 & 0 & m_{45} & m_{55} & 0 & 0 \\
 0 & 0 & 0 & 0 & 0 & m_{66} & m_{67} \\
 0 & 0 & 0 & 0 & 0 & m_{67} & m_{77} \\
\end{bmatrix}.
\end{equation}
The derived mass inertia coefficients are given in Table 2 for different face thicknesses $t_f$. The kinetic energy expression (28) will be used in the next section in the derivation of the equations of motion for the 2-D micropolar ESL-FSDT plate.
\begin{table}[ht]
\caption{Mass inertia coefficients for web-core sandwich panels modeled as 2-D micropolar ESL-FSDT plates. See Fig.~3 for the unit cell properties.}
\begin{center}
\small
\begin{tabular}{{c}|{c}{c}{c}{c}{c}}
\hline
[kg/m$^2$] & $t_f=2$ mm & $4$ mm & $6$ mm & $8$ mm & $10$ mm \\
\hline
$m_{11}$ & 37.1567 & 68.5567 & 99.9567 & 131.357 & 162.757  \\
$m_{22}$ & 37.1567 & 68.5567 & 99.9567 & 131.357 & 162.757  \\
$m_{33}$ & 37.1567 & 68.5567 & 99.9567 & 131.357 & 162.757  \\
\hline
[kg] & $t_f=2$ mm & $4$ mm & $6$ mm & $8$ mm & $10$ mm \\
\hline
$m_{44}$ & 0.016476 & 0.031692 & 0.046904 & 0.062107 & 0.077307  \\
$m_{45}$ &-0.000208 &-0.000071 &-0.000014 & 0.000007 & 0.000015  \\
$m_{55}$ & 0.000759 & 0.001776 & 0.002730 & 0.003637 & 0.004498  \\
$m_{66}$ & 0.014032 & 0.028606 & 0.043443 & 0.058422 & 0.073481  \\
$m_{67}$ & 0.000152 & 0.000345 & 0.000489 & 0.000567 & 0.000596  \\
$m_{77}$ & 0.000244 & 0.000685 & 0.001429 & 0.002399 & 0.003434  \\
\hline
\end{tabular}
\end{center}
\end{table}
\section{Governing equations and the Navier solution}
Here, we derive the equations of motion for the 2-D micropolar plate model by employing Hamilton's principle and the unit cell strain and kinetic energy expressions from the previous sections. The Navier solution will be developed for the plate in order to obtain numerical bending, buckling and natural frequency results for simply-supported web-core sandwich panels in Section 4. To study the buckling, we retain the linear constitutive matrix $\mathbf{C}$ in Eq.~(14) while the strain vector $\boldsymbol{\epsilon}$ will be augmented with von K\'arm\'an geometric nonlinearity.
\subsection{Energy equivalence}
The constitutive relation (14) was founded on energy equivalence between the macrostructure (plate) and the microstructure (unit cell). In that spirit, using the areal density $W_0^A$ of the unit cell strain energy, we write the strain energy for the micropolar plate as
\begin{equation}
U=\int_{\Omega} W_0^A\ dxdy=\frac{1}{2}\int_\Omega \boldsymbol{\epsilon}^{\textrm{T}}\mathbf{C}\boldsymbol{\epsilon}\ dxdy.
\end{equation}
In the same manner, by taking use of Eq.~(28), the total kinetic energy of the plate is
\begin{equation}
K=\int_{\Omega}\Tilde{K}_0^A\ dxdy =\frac{1}{2}\int_{\Omega}\mathbf{\dot{u}}^{\textrm{T}}\mathbf{M}\mathbf{\dot{u}}\ dxdy.
\end{equation}
The potential energy contribution due to a distributed transverse load is
\begin{equation}
V=-\int_{\Omega}qu_z\ dxdy.
\end{equation}
\subsection{Plate equilibrium equations}
By substituting expressions (30)-(32) into Hamilton's principle \citep{reddy2018}, we have
\begin{equation}
\delta\int_0^T\left[K-(U+V)\right]dt=0,
\end{equation}
which we can write in the form
\begin{equation}
\int_0^T\int_{\Omega}\left(\delta\mathbf{\dot{u}}^{\textrm{T}}\mathbf{M}\mathbf{\dot{u}}-\delta\boldsymbol{\epsilon}^{\textrm{T}}\mathbf{C}\boldsymbol{\epsilon}+q\delta u_z\right)dxdydt=0
\end{equation}
We substitute Eqs.~(14) and (23) into Eq.~(34). Moreover, in order to add the geometric nonlinearity into the 2-D micropolar plate model, we introduce the following replacements into Eq.~(34)
\begin{equation}
\begin{aligned}
N_{xx}\delta u_{x,x} &\rightarrow N_{xx}\delta\left(u_{x,x}+\underline{\frac{1}{2}u_{z,x}^2}\right), \\
N_{yy}\delta u_{y,y} &\rightarrow N_{yy}\delta\left(u_{y,y}+\underline{\frac{1}{2}u_{z,y}^2}\right), \\
N_{xy}\delta\left(u_{x,y}+u_{y,x}\right) &\rightarrow N_{xy}\delta\left[ u_{x,y}+u_{y,x}+\underline{\left(u_{z,x}u_{z,y}\right)}\right],
\end{aligned}
\end{equation}
where the underlined terms are the von K\'arm\'an strains \citep{reddy2014}. Ultimately, in this paper, the replacements (35) allow us to study linear buckling problems and good agreement will be found between 2-D micropolar and 3-D FE plate buckling results (Section 4). The explicit form of Eq.~(34) finally reads
\begin{align}
\int_0^T&\int_\Omega\Big\{m_{11}\dot{u}_x\delta\dot{u}_x+m_{22}\dot{u}_y\delta\dot{u}_y+m_{33}\dot{u}_z\delta\dot{u}_z+m_{44}\dot{\phi}_x\delta\dot{\phi}_x \notag\\
&+m_{45}\dot{\phi}_x\delta\dot{\psi}_y+m_{45}\dot{\psi}_y\delta\dot{\phi}_x+m_{55}\dot{\psi}_y\delta\dot{\psi}_y+m_{66}\dot{\phi}_y\delta\dot{\phi}_y \notag\\
&+m_{67}\dot{\psi}_x\delta\dot{\phi}_y+m_{67}\dot{\phi}_y\delta\dot{\psi}_x+m_{77}\dot{\psi}_x\delta\dot{\psi}_x+q\delta u_z\notag\\
&-\big[N_{xx}\left(\delta u_{x,x}+u_{z,x}\delta u_{z,x}\right)+N_{yy}\left(\delta u_{y,y}+u_{z,y}\delta u_{z,y}\right) \notag\\
&+N_{xy}\left(\delta u_{x,y}+\delta u_{y,x}+u_{z,x}\delta u_{z,y}+u_{z,y}\delta u_{z,x}\right) \notag\\
&+M_{xx}\delta\phi_{x,x}+M_{yy}\delta\phi_{y,y}+M_{xy}\left(\delta\phi_{y,x}+\delta\phi_{x,y}\right) \notag\\
&+Q_x^s\left(\delta\phi_x+\delta u_{z,x}\right)+Q_y^s\left(\delta\phi_y+\delta u_{z,y}\right) \notag\\
&+Q_x^a\left(\delta u_{z,x}-\delta\phi_x+2\delta\psi_y\right)+Q_y^a\left(\delta u_{z,y}-\delta\phi_y-2\delta\psi_x\right)\notag\\
&+P_{xx}\delta\psi_{x,x}+P_{yy}\delta\psi_{y,y}+P_{xy}\delta\psi_{y,x}+P_{yx}\delta\psi_{x,y}\big]\Big\}dxdydt=0.
\end{align}
We arrive at the following equations of motion (Euler--Lagrange equations) of the 2-D micropolar plate essentially by applying integration by parts in Eq.~(36):
\begin{align}
N_{xx,x}+N_{xy,y}&=m_{11}\ddot{u}_x, \\
N_{yy,y}+N_{xy,x}&=m_{22}\ddot{u}_y, \\
Q_{x,x}^s+Q_{x,x}^a+Q_{y,y}^s+Q_{y,y}^a&=m_{33}\ddot{u}_z-N-q, \\
M_{xx,x}+M_{xy,y}-Q_x^s+Q_x^a&=m_{44}\ddot{\phi}_x+m_{45}\ddot{\psi}_y, \\
P_{yy,y}+P_{xy,x}-2Q_x^a&=m_{55}\ddot{\psi}_y+m_{45}\ddot{\phi}_x, \\
M_{yy,y}+M_{xy,x}-Q_y^s+Q_y^a&=m_{66}\ddot{\phi}_y+m_{67}\ddot{\psi}_x, \\
P_{xx,x}+P_{yx,y}+2Q_y^a&=m_{77}\ddot{\psi}_x+m_{67}\ddot{\phi}_y,
\end{align}
where
\begin{equation}
N=\frac{\partial}{\partial x}\left(N_{xx}u_{z,x}+N_{xy}u_{z,y}\right)+\frac{\partial}{\partial y}\left(N_{yy}u_{z,y}+N_{xy}u_{z,x}\right).
\end{equation}
It can be seen explicitly from Eqs.~(41) and (43) how the antisymmetric shear behavior is coupled to the local bending and twisting of the plate, as was discussed already in Section 1. We define the following shear resultants
\begin{equation}
\begin{aligned}
Q_{xz}&\equiv Q_{x}^s+Q_{x}^a, \quad Q_{zx}\equiv Q_{x}^s-Q_{x}^a, \\
Q_{yz}&\equiv Q_{y}^s+Q_{y}^a, \quad Q_{zy}\equiv Q_{y}^s-Q_{y}^a.
\end{aligned}
\end{equation}
Figure 4 shows how $Q_{xz}$ and $Q_{zx}$ are formed from the symmetric and antisymmetric parts $Q_{x}^s$ and $Q_{x}^a$, respectively, or in other words, how $Q_{xz}$ and $Q_{zx}$ are split into symmetric and antisymmetric parts. The definitions (45) correlate with the shear strains (3) and (4), e.g., $\gamma_{xz}\equiv\gamma_{x}^s+\gamma_{x}^a=2\epsilon_{xz}$. Now, if the dynamic effects are ignored and $N=0$ and we use Eqs.~(45), the equilibrium equations (37)-(43) are the same (apart from the face loads) as those obtained by integrating the stresses acting on an infinitesimal parallelepiped micropolar 3-D element with respect to the thickness coordinate $z$, see Appendix C. That is to say, by using Eqs.~(45) the derived constitutive relations (15) could be plugged directly into the general equilibrium equations (C.2) based on 3-D micropolar elasticity to study static, geometrically linear 2-D web-core plate bending problems.
\begin{figure}
\centering
\includegraphics[scale=1]{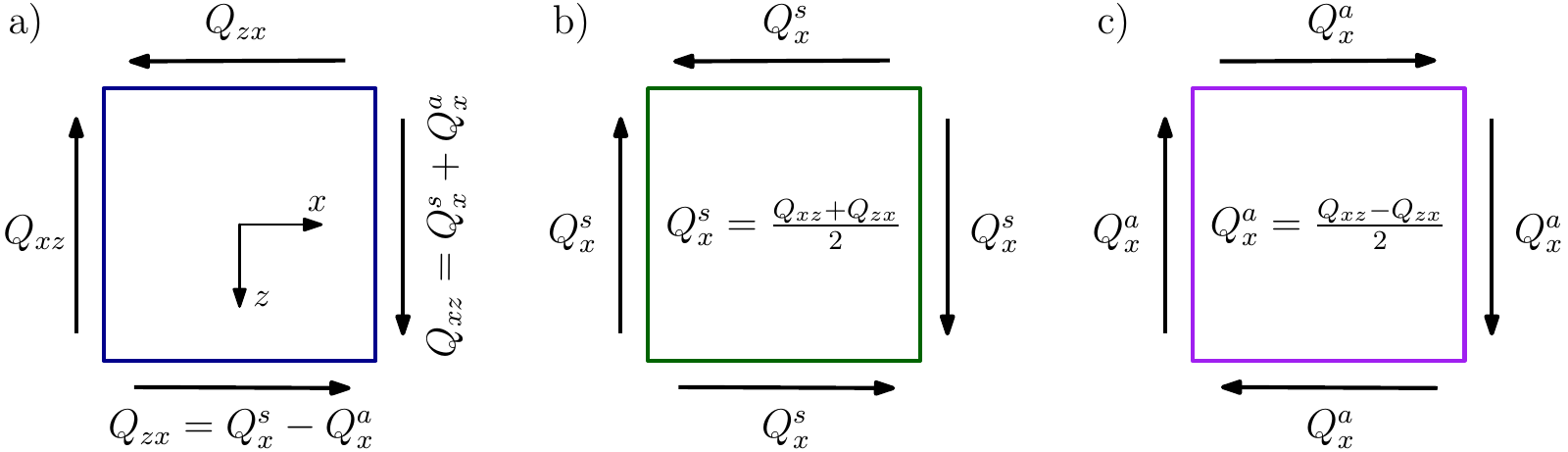}
\caption{a) Split of the shear resultants $Q_{xz}$ and $Q_{zx}$ into b) symmetric and c) antisymmetric parts, see \citep{mindlin1963}}
\end{figure}
\subsection{Plate boundary conditions}
Figure 5(a) shows the stress resultants acting on the edges of the 2-D micropolar plate. Figure 5(b) shows the setup for the Navier solution of a rectangular simply-supported plate. In general, the boundary conditions for edges $x=(0,a)$ of the rectangular plate in Fig.~5(b) are set by specifying one element in each of the following pairs:
\begin{equation}
\begin{aligned}
&N_{xx} \ \ \textrm{or} \ \ u_x, \quad N_{xy} \ \ \textrm{or} \ \ u_y, \quad (Q_{xz}+N_{xx}u_{z,x}+N_{xy}u_{z,y}) \ \ \textrm{or} \ \ u_z \\
&M_{xx} \ \ \textrm{or} \ \ \phi_x, \quad M_{xy} \ \ \textrm{or} \ \ \phi_y, \quad P_{xx} \ \ \textrm{or} \ \ \psi_x, \quad P_{xy} \ \ \textrm{or} \ \ \psi_y
\end{aligned}
\end{equation}
and on edges $y=(0,b)$ we impose
\begin{equation}
\begin{aligned}
&N_{yy} \ \ \textrm{or} \ \ u_y, \quad N_{xy} \ \ \textrm{or} \ \ u_x, \quad (Q_{yz}+N_{yy}u_{z,y}+N_{xy}u_{z,x}) \ \ \textrm{or} \ \ u_z \\
&M_{yy} \ \ \textrm{or} \ \ \phi_y, \quad M_{xy} \ \ \textrm{or} \ \ \phi_x, \quad P_{yy} \ \ \textrm{or} \ \ \psi_y, \quad P_{yx} \ \ \textrm{or} \ \ \psi_x.
\end{aligned}
\end{equation}
Only the bending behavior will be considered here in the Navier solution as reflected by the boundary conditions given in Fig.~5(b).

\medskip
\begin{figure}
\centering
\includegraphics[scale=0.9]{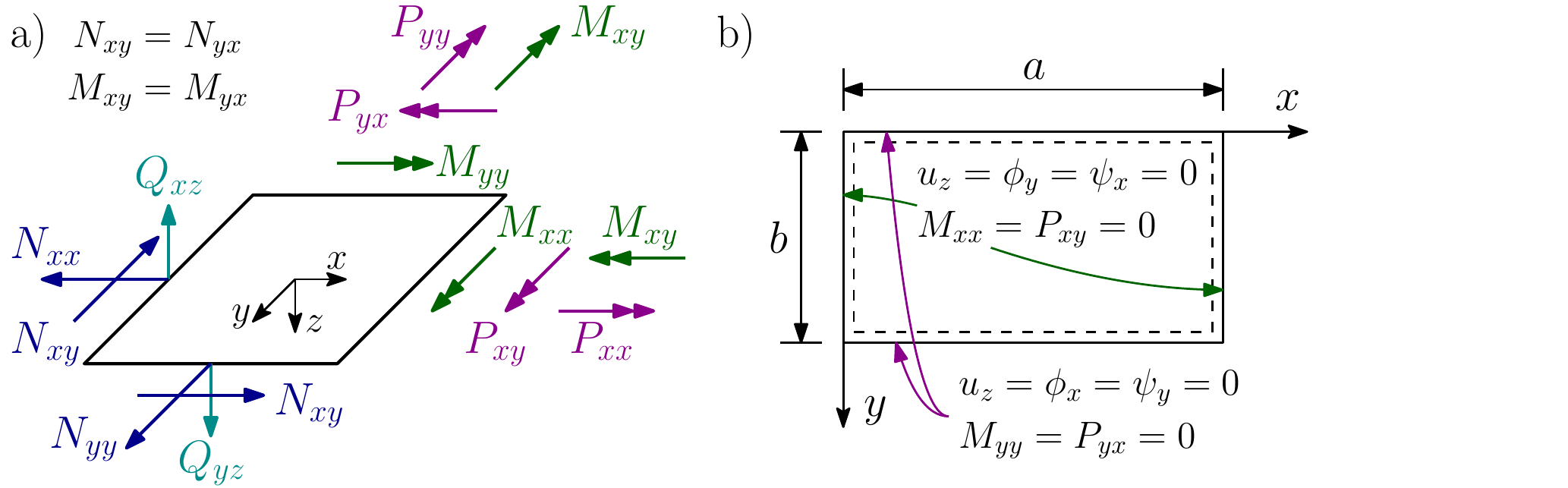}
\caption{a) Stress resultants acting on the edges of a 2-D micropolar ESL-FSDT plate. The positive directions are shown and the directions are reversed for opposite faces. b) Boundary conditions for the bending of a rectangular simply-supported plate.}
\end{figure}

\subsection{The Navier solution for bending, buckling, and free vibration}
The simply-supported boundary conditions displayed in Fig.~5(b) are satisfied by the following choices of displacement and load expansions:
\begin{equation}
\begin{aligned}
\begin{Bmatrix}
u_z \\
q
\end{Bmatrix}
&=\sum_{n=1}^{\infty}\sum_{m=1}^{\infty}
\begin{Bmatrix}
W_{mn}(t) \\
Q_{mn}(t)
\end{Bmatrix}
\sin\alpha x\sin\beta y, \\
\begin{Bmatrix}
\phi_x \\
\psi_y
\end{Bmatrix}
&=\sum_{n=1}^{\infty}\sum_{m=1}^{\infty}
\begin{Bmatrix}
X_{mn}(t) \\
P_{mn}(t)
\end{Bmatrix}
\cos\alpha x\sin\beta y, \\
\begin{Bmatrix}
\phi_y \\
\psi_x
\end{Bmatrix}
&=\sum_{n=1}^{\infty}\sum_{m=1}^{\infty}
\begin{Bmatrix}
Y_{mn}(t) \\
R_{mn}(t)
\end{Bmatrix}
\sin\alpha x\cos\beta y,
\end{aligned}
\end{equation}
where
\begin{equation}
\begin{aligned}
\alpha=\frac{m\pi}{a} \quad \textrm{and} \quad \beta=\frac{n\pi}{b},
\end{aligned}
\end{equation}
and
\begin{equation}
Q_{mn}(t)=\frac{4}{ab}\int_0^b\int_0^a q\sin\alpha x\sin\beta y \, dxdy.
\end{equation}
Substitution of Eqs.~(48) into Eqs.~(39)-(43) yields the following equation for the coefficients $(W_{mn},X_{mn},P_{mn},Y_{mn},R_{mn})$
\begin{equation}
\hat{\mathbf{M}}\ddot{\mathbf{w}}+\hat{\mathbf{K}}\mathbf{w}
=\mathbf{Q},
\end{equation}
where
\begin{equation}
\begin{aligned}
&\mathbf{w}=\left\{W_{mn} \ \ X_{mn} \ \ P_{mn} \ \ Y_{mn} \ \ R_{mn}\right\}^{\textrm{T}}, \\
&\mathbf{Q}=\left\{Q_{mn} \ \ 0 \ \ 0 \ \ 0 \ \ 0 \right\}^{\textrm{T}},
\end{aligned}
\end{equation}
and
\begin{equation}
\hat{\mathbf{K}}=
\begin{bmatrix}
\hat{k}_{11} & \alpha(G_{11}-G_{22}) & 2\alpha(G_{12}+G_{22}) & \beta(G_{33}-G_{44}) & -2\beta(G_{34}+G_{44}) \\
 & \hat{k}_{22} & 2(G_{12}-G_{22}) & \alpha\beta(D_{12}+D_{33}) & 0 \\
 & & \hat{k}_{33} & 0 & \alpha\beta(H_{12}+H_{34}) \\
 & \textrm{SYM} & & \hat{k}_{44} & 2(G_{44}-G_{34}) \\
 & & & & \hat{k}_{55}
\end{bmatrix}
\end{equation}
\begin{equation}
\begin{aligned}
\hat{k}_{11}&=\alpha^2(G_{11}+2G_{12}+G_{22}-kN_0)+\beta^2(G_{33}+2G_{34}+G_{44}-N_0), \\
\hat{k}_{22}&=G_{11}-2G_{12}+G_{22}+\alpha^2 D_{11}+\beta^2 D_{33}, \\
\hat{k}_{44}&=G_{33}-2G_{34}+G_{44}+\beta^2 D_{22}+\alpha^2 D_{33}, \\
\hat{k}_{33}&=4G_{22}+\beta^2 H_{22}+\alpha^2 H_{33}, \hat{k}_{55}=4G_{44}+\alpha^2 H_{11}+\beta^2 H_{44}.
\end{aligned}
\end{equation}
In Eqs.~(54), the following in-plane biaxial compressive forces are used in buckling analysis:
\begin{equation}
\hat{N}_{xx}=-kN_0, \quad \hat{N}_{yy}=-N_0, \quad k=\frac{\hat{N}_{xx}}{\hat{N}_{yy}} \quad (N_{xy}=0).
\end{equation}
Matrix $\hat{\mathbf{M}}$ is of the same form as that given by Eq.~(29) with the exception that now $m_{11}=m_{22}=0$. For further details on how Eq.~(51) is used to calculate actual results, see the book by \cite{reddy2004}.
\section{Numerical examples}
Here, we first study the static bending of simply-supported web-core sandwich panels under line and uniformly distributed loads. Second, the buckling of web-core panels subjected to uniaxial and biaxial compression is investigated. Finally, the natural vibration frequencies of a web-core panel are calculated. The numerical studies are carried out using the Navier solution for the two-dimensional ESL-FSDT plate models based on micropolar and classical elasticity. In addition, a 3-D finite element model for web-core sandwich panels is developed to provide accurate reference solutions against which the ESL-FSDT models can be evaluated.
\begin{figure}[hb]
\centering
\includegraphics[scale=0.59]{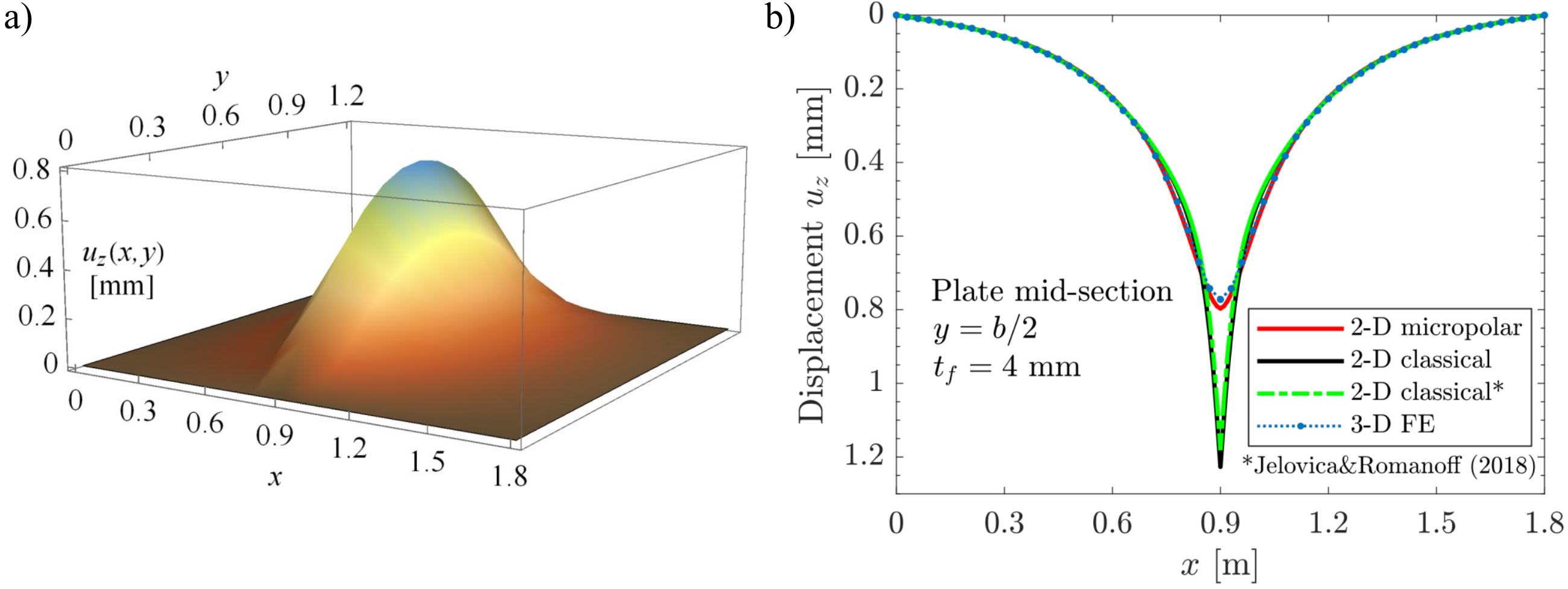}
\caption{a) Transverse displacement of a 2-D micropolar ESL-FSDT web-core plate under a line load for $t_f=4$ mm (see Fig.~1 for 3-D results). b) Transverse displacement of the plate mid-section calculated by different methods.}
\end{figure}
\subsection{3-D FE reference model}
A 3-D finite element web-core panel modeled using Abaqus 2018 was displayed in Fig.~1. In all the following analyses the global element size of the model is 8 mm, which yields convergent results. Three different size plates will be studied; the plate planform area is $(a\cdot b)$ m$^2$ and the studied sizes are $(1.2\cdot 1.2)$ m$^2$, $(1.8\cdot 1.2)$ m$^2$, and $(2.04\cdot 2.04)$ m$^2$, and the corresponding 3-D FE models consist of 60800, 91200, and 156570 shell elements of type S8R5, respectively. The boundary conditions are imposed in a similar manner as in classical simply-supported 3-D solid plate problems. For all nodes $i=1,2,\ldots,n$ on edges $x=(0,a)$ (see Fig.~5) we use $U_z^i=U_y^i=\Psi_x^i=0$. Analogously, for all nodes on edges $y=(0,b)$ we use $U_z^i=U_x^i=\Psi_y^i=0$. In the Abaqus buckling analysis, the boundary conditions are applied at two stages. For ``Stress perturbation" one needs to set on edges $x=(0,a)$ the conditions $U_z^i=\Psi_x^i=0$ and for ``Buckling mode calculation" $U_z^i=U_y^i=\Psi_x^i=0$ are to be used. A corresponding approach is used for edges $y=(0,b)$.
\subsection{Bending analysis}
Let us consider a $(1.8\cdot 1.2)$ m$^2$ panel with the center web subjected to a line load so that (Fig.~1)
\begin{equation}
Q_{mn}=\frac{8q_0}{\pi na}\sin\frac{m\pi}{2} \quad (m=1,2,3,\ldots;n=1,3,5,\ldots)
\end{equation}
where $q_0=10000$ N/m. The static bending solution can be obtained from Eq.~(51) by setting the time derivative terms and edge forces $(\hat{N}_{xx}, \hat{N}_{yy})$ to zero.
\begin{figure}[hb]
\centering
\includegraphics[scale=0.95]{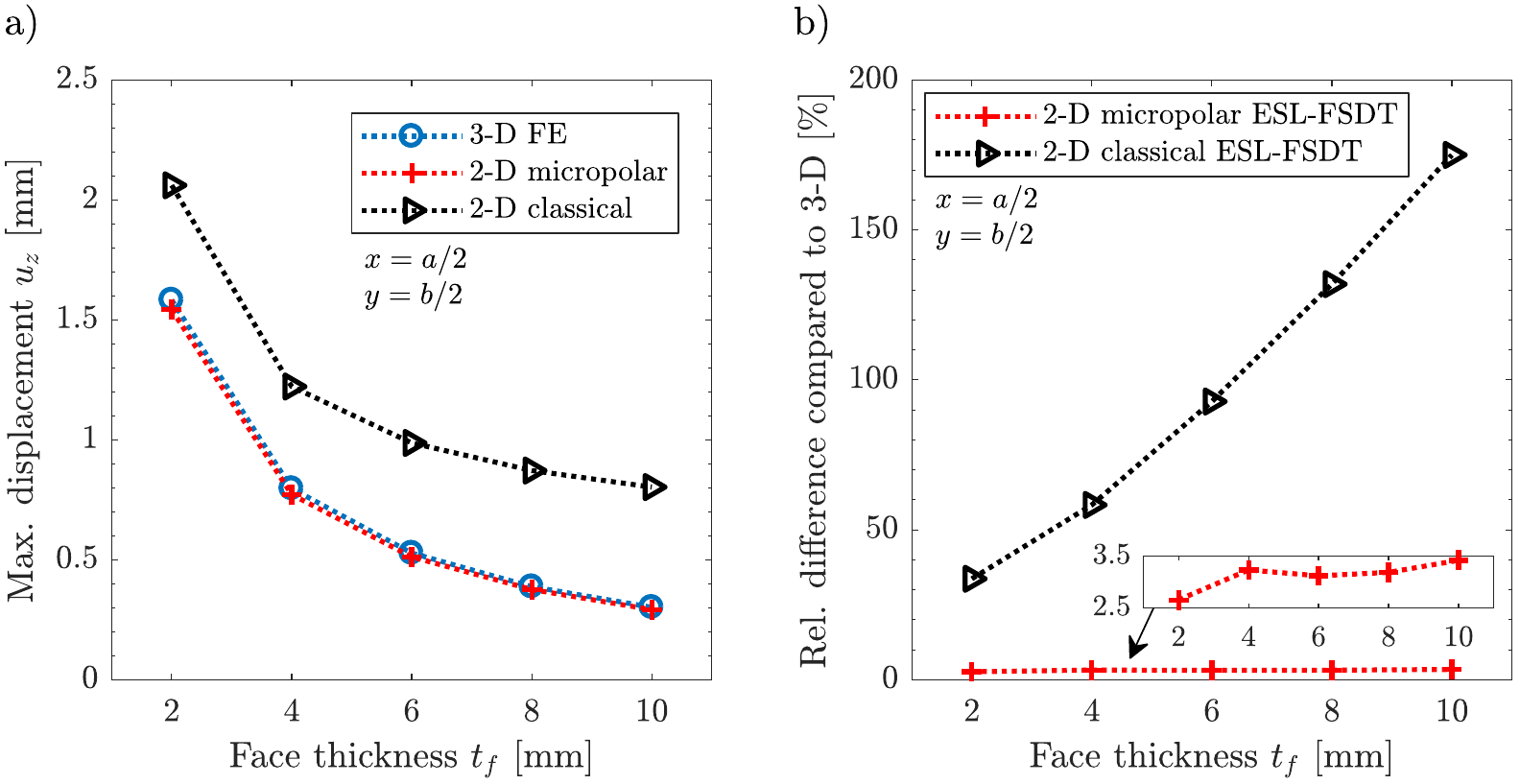}
\caption{a) Maximum transverse displacement of a web-core sandwich panel under a line load calculated by a 3-D FE model and 2-D ESL-FSDT models based on micropolar and classical elasticity. b) Relative difference between the 2-D and 3-D models in terms of the maximum displacement.}
\end{figure}

Figure 6(a) shows the transverse displacement of the 2-D micropolar ESL-FSDT plate. Figure 6(b) shows the displacement at the plate mid-section $y=b/2$ as given by different models. The 2-D micropolar plate results and those by the 3-D FE model are in good agreement, whereas the 2-D ESL-FSDT plate model based on the classical elasticity theory performs poorly near the center of the plate. Two different sets of constitutive parameters are used in the classical case; those from Table B.3 and the quite differently defined parameters used recently by \cite{jelovica2018}. The classical model cannot capture the antisymmetric shear behavior nor the local face bending that occurs near the line load (see Fig.~1); thus, a large overshoot occurs in the displacement there when the 2-D classical ESL-FSDT plate model is used. Figure 7(a) shows the maximum transverse displacements at the plate center for face thicknesses $t_f=2,4,\ldots 10$ mm and Fig.~7(b) shows the relative difference
\begin{equation}
\Delta u_z=100\cdot\frac{u_{z,\textrm{2-D}}-u_{z,\textrm{3-D}}}{u_{z,\textrm{3-D}}}\ [\%]
\end{equation}
between the 2-D ESL-FSDT and the 3-D FE models. It can be seen in Fig.~7(b) that as the face thickness $t_f$ increases, the difference between the classical 2-D and 3-D FE model grows since an increase in $t_f$ amplifies the roles of the antisymmetric shear and local bending and twisting. For example, Eq.~(26) shows that the local bending stiffness $H_{33}$ is proportional to $t_f^3$. In the case of the classical plate model, we have $\Delta u_z=34-175\%$ for $t_f=2-10$ mm, whereas for the micropolar plate model, Eq.~(57) gives an acceptable range of $\Delta u_z=2.7-3.4\%$.

Figure 8(a) displays the mid-section transverse displacement of a square $(1.2\cdot1.2)$ m$^2$ plate under a uniform load $Q_{mn}=16q_0/(\pi mn)$, where $q_0=10000$ N/m$^2$ $(m,n=1,3,5,\ldots)$. The 2-D micropolar and 3-D FE results are in good agreement. In this case the 2-D classical result is also satisfactory as the uniform load does not induce the kind of deformations like a line or point load that would require the use of the micropolar model. Nevertheless, Fig.~8(b) shows that when the face thickness $t_f$ increases, the relative difference increases at a faster rate with the classical model.

\begin{figure}[ht]
\centering
\includegraphics[scale=0.95]{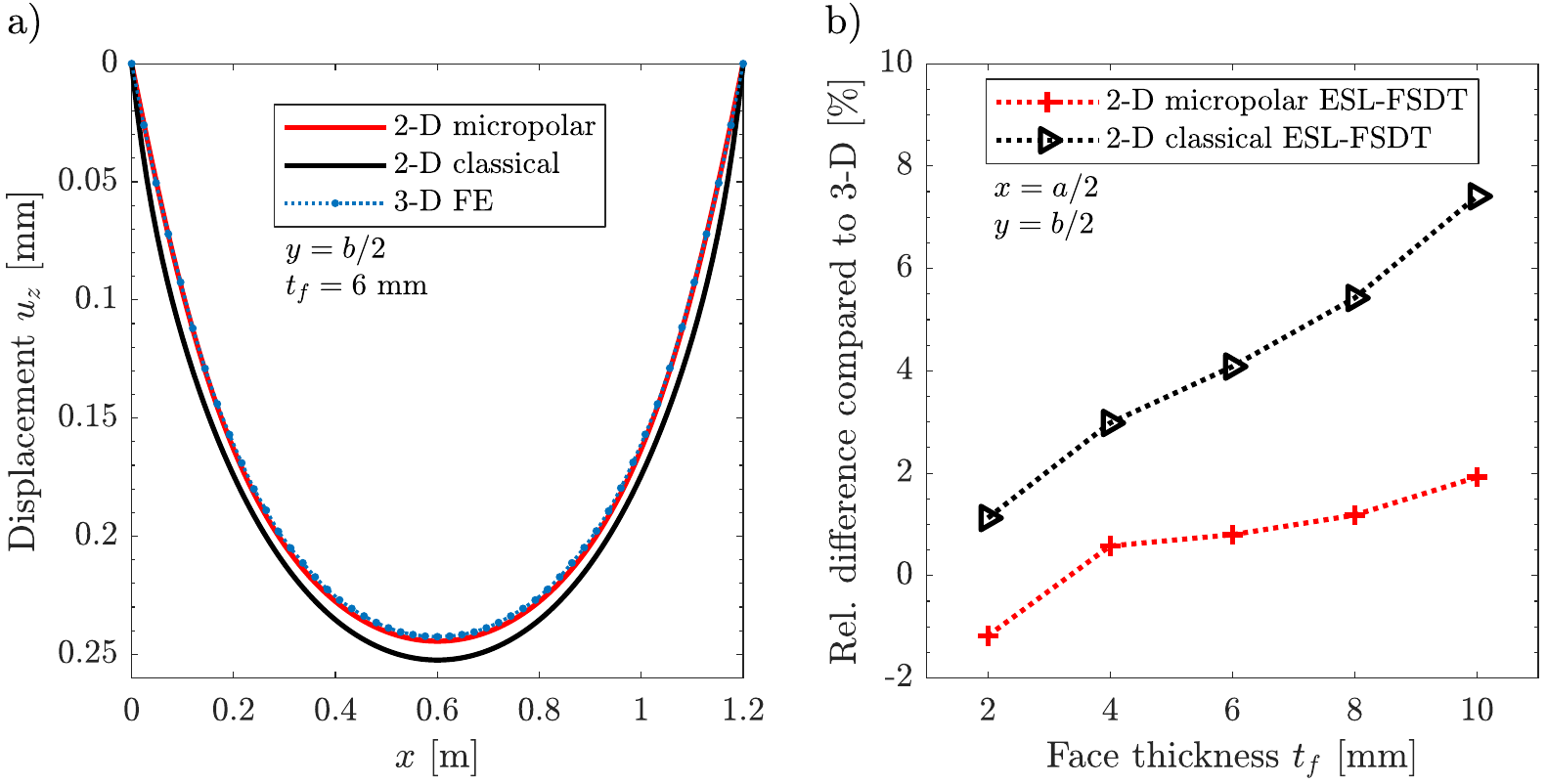}
\caption{a) Transverse displacement of the mid-section a web-core sandwich panel under a uniform distributed load calculated by a 3-D FE model and 2-D ESL-FSDT models based on micropolar and classical elasticity. b) Relative difference between the 2-D and 3-D models in terms of the maximum displacement.}
\end{figure}
\begin{figure}
\centering
\includegraphics[scale=0.6]{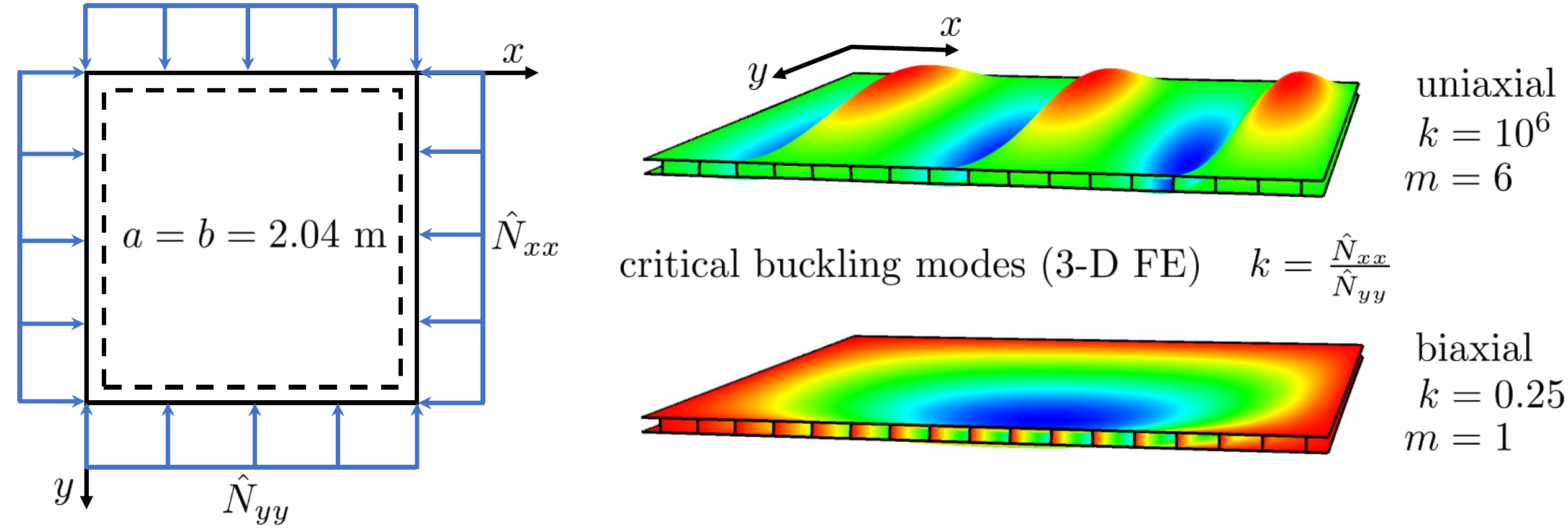}
\caption{Setup for web-core plate buckling analysis and the critical buckling modes for uniaxial and biaxial compression.}
\end{figure}
\subsection{Uniaxial and biaxial buckling}
For the buckling analysis, we set the time derivative terms and the transverse load to zero in Eq.~(51). Figure 9 presents the setup for the buckling problems and the critical buckling modes calculated by the 3-D FE model. Figures 10(a) and 10(b) show the lowest buckling loads up to $m=11$ (for all of which $n=1$) under uniaxial and biaxial compression, respectively. In both cases we can see that the 2-D micropolar ESL-FSDT plate results are in good agreement with the 3-D FE results, whereas the 2-D classical results are not. The classical model can predict well only the lowest mode ($m=1, n=1$) corresponding to the deformation pattern under the uniform load [Figs.~8(a) and 9]. However, the classical model predicts the critical mode incorrectly in both cases, as the buckling loads decrease in an non-physical manner as the mode number increases.
\begin{figure}[h!]
\centering
\includegraphics[scale=0.95]{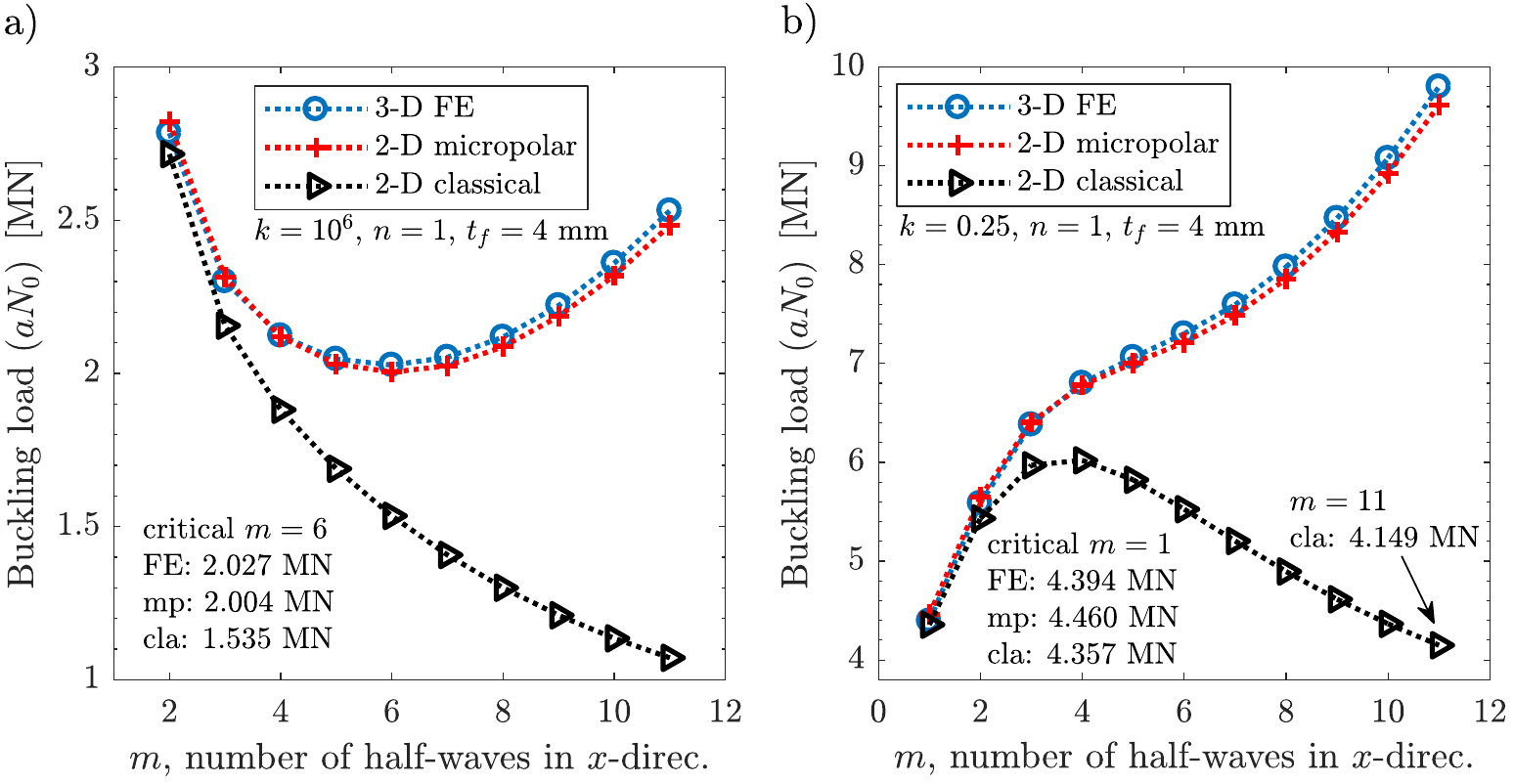}
\caption{Buckling loads of a web-core sandwich panel under a) uniaxial and b) biaxial compression. The ESL-FSDT plate model based on classical elasticity gives erroneous results for the critical buckling mode and load, whereas the 2-D micropolar model is in good agreement with the 3-D results in both cases.}
\end{figure}

\subsection{Natural vibration frequencies}
For free vibration, we consider a $(1.8\cdot1.2)$ m$^2$ plate and set the mechanical loads to zero and seek a periodic solution to Eq.~(51) in the form $\mathbf{w}=\mathbf{w}^0e^{i\omega t}$. The eight lowest natural frequencies for the 2-D and 3-D plate models are presented in Fig.~11. Both 2-D plate models provide accurate estimates for the fundamental vibration frequency $f_{1,1}$. However, as the mode number increases, the 2-D classical ESL-FSDT plate model becomes essentially too flexible and underpredicts the natural frequencies. This behavior can be problematic in particular when the forced response of a web-core sandwich panel is calculated by taking use of mode superposition.
\begin{figure}
\centering
\includegraphics[scale=1]{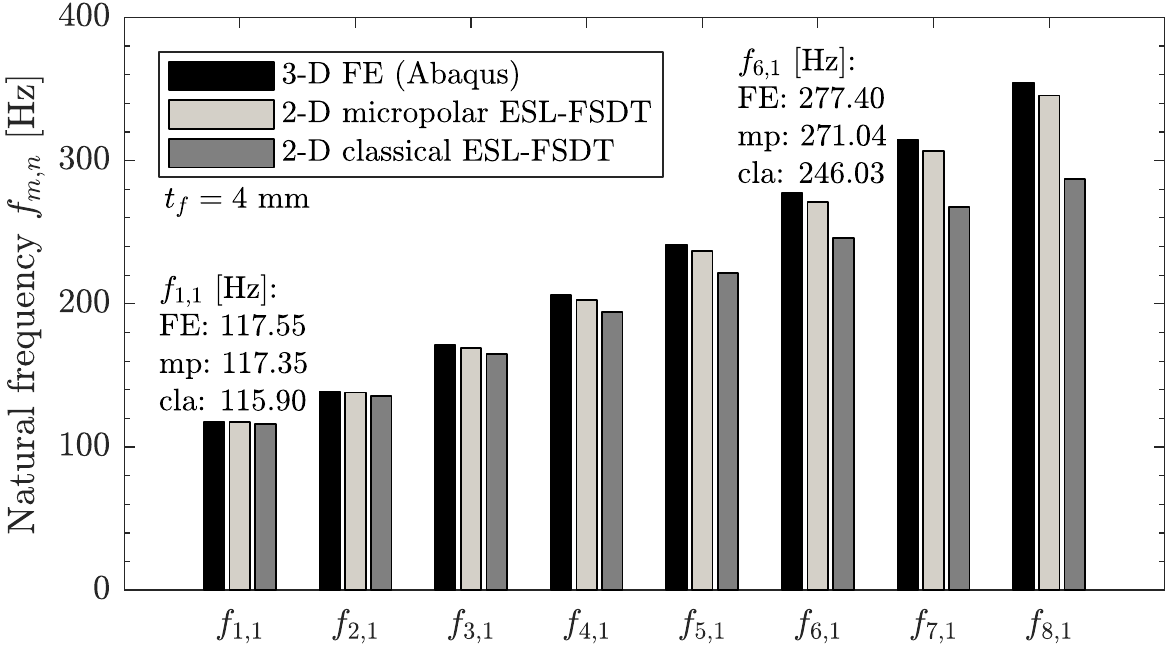}
\caption{Eight lowest natural vibration frequencies of a web-core sandwich panel calculated by a 3-D FE model and 2-D ESL-FSDT models based on micropolar and classical elasticity. As the mode number increases, the classical ESL plate model becomes inaccurate which can be problematic if mode superposition is applied in response analysis.}
\end{figure}
\section{Concluding remarks}
A novel 2-D micropolar equivalent single-layer (ESL) plate model for orthotropic web-core sandwich panels was introduced. The plate allows for arbitrary but constant rotation of the plate transverse normals like the conventional first-order shear deformation plate theory (FSDT) based on the classical elasticity. The derivation of the 2-D micropolar ESL-FSDT plate model was built upon two-scale modeling of a 3-D web-core unit cell that gave, through energy equivalence between the unit cell and the plate, the strain and kinetic energy expressions for the 2-D plate. These were then used to derive the constitutive relations and the equations of motion for the plate.

The micropolar approach was used here because it allowed the passage of information on both displacements and rotations from the discrete 3-D web-core unit cell modeled by classical shell finite elements into the 2-D ESL plate continuum. This way it was possible to fully account for the deformations of a 3-D panel in a dimensionally-reduced 2-D ESL plate continuum through non-classical antisymmetric shear behavior and local bending and twisting.

The 2-D micropolar plate model provides homogenized, computationally cost-effective solutions to bending, buckling, and free vibration problems of web-core sandwich panels used, for example, in large-scale ship and bridge decks. The small differences between the presented numerical 2-D micropolar and 3-D FE results may be due to the boundary conditions; in 2-D problems they are imposed only at the mid-surface, whereas in 3-D they are enforced practically throughout the plate thickness. Thus, in 3-D the plate tends to be behave in a bit stiffer manner. This issue has been studied in detail recently in the context of classical elasticity \citep{karttunen2018c}. It is important to note that at this stage the presented 2-D micropolar cannot capture higher, local vibration and buckling modes that may appear within a unit cell at higher frequencies and loads than were studied here. For additional free vibration and buckling analyses of web-core panels by classical and 3-D FE models, see also the papers by \cite{jelovica2016} and \cite{jelovica2018}.

The presented micropolar two-scale energy approach for constitutive modeling was constructed so that it is independent of the core of the unit cell and, thus, has good potential for extensions to other lattice core topologies. The approach may also be applicable in the context of other non-classical continuum mechanics theories [see, e.g. \citep{khakalo2018}]. We note that there are different strategies to obtain the periodic, classical 3-D stresses of a web-core sandwich panel by post-processing the 2-D micropolar results. For web-core beams this was done in \citep{karttunen2019a} and for full 3-D web-core panels this will be carried out in future studies. Finally, this study introduced a theoretical micropolar plate model which can be used as the basis for micropolar finite element developments similar to those presented by \cite{ansari2017,ansari2018}. Furthermore, the micromorphic continuum theory may provide a suitable framework for modeling local buckling and vibrations of lattice core sandwich panels through its additional degrees of freedom \citep{eringen2012,isbuga2011,ansari2016,ansari2017b}.

\section*{Acknowledgements}
The first author acknowledges that this work has received funding from the European Union's Horizon 2020 research and innovation programme under the Marie Sk\l{}odowska--Curie grant agreement No 745770. The financial support is greatly appreciated. The authors also wish to acknowledge CSC -- IT Center for Science, Finland, for computational resources (Abaqus usage).
\appendix
\section{Unit cell modeling}
The finite element unit cell in Fig.~2 is modeled using Abaqus thin shell elements S4R5. However, in practice, we use S8R5 thin shell elements with mid-side nodes in this paper. The mid-side nodes are treated in the same way as
element corner nodes when the discrete-to-continuum transformation (7) is applied along the face edges. The S8R5 elements have five degrees of freedom at each node including three translations and two rotations but there is no drilling degree of freedom associated with the membrane part of the element. Thus, the condensed-out global drilling of the unit cell in Fig.~2 is mainly related to the bending of the web with respect to the global $z$-axis.

In Abaqus, the static condensation (Guyan reduction) of the unit cell is carried out in a ``Substructure generation" step. After mesh generation, the face edge nodes $i$ are renumbered so that $i=1,2,\ldots,n$. In order to obtain the reduced stiffness and mass matrices, the command ``*SUBSTRUCTURE MATRIX OUTPUT, OUTPUT FILE=USER DEFINED, FILE NAME=xx, STIFFNESS=YES, MASS=YES" has to be added to the input file manually. The Abaqus output matrices can be read in Matlab by using the readily available ``leerSubs.m" script (\textit{Matlab code to process Abaqus output}) found at https://github.com/Ondiz/MatlabAbaqus. In Matlab, it is then straightforward to create a discrete-to-continuum transformation script based on Eqs.~(6). The script should take the reduced unit cell stiffness and mass matrices as well as the nodal coordinate data (from Abaqus .inp file) as input data. Equations (6) are then applied at nodes $i=1,2,\ldots,n$ and the Matlab command 'equationsToMatrix' can be used to construct the necessary transformation matrices. It is healthy to add a numerical check to the end of the continualization script to see that Eq.~(11) holds, i.e., the displacement vector $\mathbf{u}$ should not contribute to the strain energy.

The required unit cell mesh size can be determined by considering the convergence of the constitutive matrix (13). With a global element size of 3 mm, the unit cell consists of 3800 shell elements (S8R5), and a size of 2 mm leads to a total of 8520 elements. The relative differences of all components of both $\mathbf{D}$ and $\mathbf{H}$ matrices, which consider two different types of bending and twisting, between the two different mesh sizes are very close to 0\%. A global mesh size of 2 mm is used in all unit cell analyses in this paper. Each unit cell needs to be analysed only once to obtain the constitutive and mass inertia coefficients and the unit cell mesh size does not relate to any critical computational efficiency issues in the present context because the FE unit cells as such are not used to solve actual 2-D plate problems.
\section{Classical constitutive modeling}
The constitutive modeling approach, as outlined in Section 2, can also be used in the context of classical elasticity, and it is a novel contribution in this respect as well. In this case, all rotational degrees of freedom of a FE unit cell modeled by shell finite elements are condensed out and only phase (1) of Section 2.3 needs to be carried out to obtain the constitutive matrix. All variables and coefficients in this Appendix relate to conventional ESL-FSDT plates \citep{reddy2004} and classical elasticity regardless of the fact that the used notations overlap with the micropolar ones; The treatment by Eqs.~(1)--(55) is for the micropolar modeling approach. For a 2-D conventional ESL-FSDT plate the displacements are
\begin{equation}
U_x=u_x+z\phi_x, \quad U_y=u_y+z\phi_y, \quad U_z=u_z
\end{equation}
and the nonzero strains of the plate are
\begin{equation}
\begin{aligned}
\epsilon_{xx}&=U_{x,x}=u_{x,x}+z\phi_{x,x}=\epsilon_{xx}^0+z\kappa_{xx}, \\
\epsilon_{yy}&=U_{y,y}=u_{y,y}+z\phi_{y,y}=\epsilon_{yy}^0+z\kappa_{yy}, \\
\gamma_{xy}&=U_{x,y}+U_{y,x}=u_{x,y}+u_{y,x}+z\left(\phi_{x,y}+\phi_{y,x}\right)\\
&=\epsilon_{yx}^0+\epsilon_{xy}^0+z\left(\kappa_{yx}+\kappa_{xy}\right), \\
\gamma_{xz}&=U_{z,x}+U_{x,z}=u_{z,x}+\phi_x, \\
\gamma_{yz}&=U_{z,y}+U_{y,z}=u_{z,y}+\phi_y .
\end{aligned}
\end{equation}
With distance from an arbitrary fiber located within a plate domain (Fig.~2), Taylor series expansion of Eqs.~(B.1) for edge nodes $i=1,2,\ldots,n$ on the bottom and top faces $(z=\pm h/2)$ leads to
\begin{equation}
\begin{aligned}
U_x^i&=u_x\pm\frac{h}{2}\left(\frac{\gamma_{xz}}{2}+\omega_{y}\right)+\Delta x^i\left[\epsilon^0_{xx}\pm\frac{h}{2}\kappa_{xx}\right]+\Delta y^i\left[\epsilon^0_{yx}\pm\frac{h}{2}\kappa_{yx}\right], \\
U_y^i&=u_y\pm\frac{h}{2}\left(\frac{\gamma_{yz}}{2}-\omega_{x}\right)+\Delta y^i\left[\epsilon^0_{yy}\pm\frac{h}{2}\kappa_{yy}\right]+\Delta x^i\left[\epsilon^0_{xy}\pm\frac{h}{2}\kappa_{xy}\right], \\
U_z^i&=u_z+\Delta x^i\left(\frac{\gamma_{xz}}{2}-\omega_{y}\right)+\Delta y^i\left(\frac{\gamma_{yz}}{2}+\omega_{x}\right)
\end{aligned}
\end{equation}
where $\Delta x^i$ and $\Delta y^i$ are the nodal coordinates of the FE unit cell. In addition, $(\omega_x,\omega_y)$ are the macrorotations, see Eqs.~(5). A matrix formulation analogous to Eqs.~(7)-(15) will lead to the constitutive membrane $\mathbf{A}$, bending and twisting $\mathbf{D}$ and shear $\mathbf{D}_Q$ matrices
\begin{align}
\mathbf{A}=
\begin{bmatrix}
 A_{11} & A_{12} & 0  \\
 A_{12} & A_{22} & 0 \\
 0 & 0 & A_{33}
\end{bmatrix},
\:
\mathbf{D}=
\begin{bmatrix}
 D_{11} & D_{12} & 0  \\
 D_{12} & D_{22} & 0  \\
 0 & 0 & D_{33}
\end{bmatrix},
\:
\mathbf{D}_Q=
\begin{bmatrix}
D_{Qx} & 0  \\
0 & D_{Qy}
\end{bmatrix}.
\end{align}
The conventional constitutive parameters are listed in Table B.3. Extrinsic shear corrections factors are not employed here.

\begin{table}[ht]
\caption{Constitutive coefficients for web-core sandwich panels modeled as 2-D conventional ESL-FSDT plates. See Fig.~3 for the unit cell properties.}
\begin{center}
\small
\begin{tabular}{{c}|{c}{c}{c}{c}{c}}
\hline
$\mathbf{A}$ [MN/m] & $t_f=2$ mm & $4$ mm & $6$ mm & $8$ mm & $10$ mm \\
\hline
$A_{11}$ & 905.495 & 1810.99 & 2716.48 & 3621.98 & 4527.47 \\
$A_{12}$ & 271.648 & 543.297 & 814.945 & 1086.59 & 1358.24 \\
$A_{22}$ & 997.007 & 1907.02 & 2814.27 & 3720.74 & 4626.24 \\
$A_{33}$ & 316.923 & 633.846 & 950.769 & 1267.69 & 1584.62 \\
\hline
$\mathbf{D}$ [MNm] & $t_f=2$ mm & $4$ mm & $6$ mm & $8$ mm & $10$ mm \\
\hline
$D_{11}$ & 0.43826 & 0.87652 & 1.31478 & 1.75304 & 2.19130 \\
$D_{12}$ & 0.13148 & 0.26296 & 0.39443 & 0.52591 & 0.65739 \\
$D_{22}$ & 0.45844 & 0.90014 & 1.34456 & 1.79037 & 2.23576 \\
$D_{33}$ & 0.15339 & 0.30678 & 0.46017 & 0.61356 & 0.76695 \\
\hline
$\mathbf{D}_Q$ [MN/m] & $t_f=2$ mm & $4$ mm & $6$ mm & $8$ mm & $10$ mm \\
\hline
$D_{Qx}$ & 0.18414 & 0.30338 & 0.32673 & 0.33315 & 0.33556 \\
$D_{Qy}$ & 31.4205 & 38.2281 & 41.8695 & 44.0560 & 45.4626 \\
\hline
\end{tabular}
\end{center}
\end{table}
The conventional mass inertia coefficients are obtained following the approach taken in Section 2.4 and are given in Table B.4. The parameters form a diagonal matrix $[m_{11},m_{22},m_{33},m_{44},m_{66}]$ that corresponds to a velocity vector $[\dot{u}_x,\dot{u}_y,\dot{u}_z,\dot{\phi}_x,\dot{\phi}_y]$, see Eq.~(28) for reference.
\begin{table}[ht]
\caption{Mass inertia coefficients for web-core sandwich panels modeled as 2-D conventional ESL-FSDT plates. See Fig.~3 for the unit cell properties.}
\begin{center}
\small
\begin{tabular}{{c}|{c}{c}{c}{c}{c}}
\hline
[kg/m$^2$] & $t_f=2$ mm & $4$ mm & $6$ mm & $8$ mm & $10$ mm \\
\hline
$m_{11}$ & 37.1567 & 68.5567 & 99.9567 & 131.357 & 162.757  \\
$m_{22}$ & 37.1567 & 68.5567 & 99.9567 & 131.357 & 162.757  \\
$m_{33}$ & 37.1567 & 68.5567 & 99.9567 & 131.357 & 162.757  \\
\hline
[kg] & $t_f=2$ mm & $4$ mm & $6$ mm & $8$ mm & $10$ mm \\
\hline
$m_{44}$ & 0.01665 & 0.03171 & 0.04690 & 0.06211 & 0.07731  \\
$m_{66}$ & 0.01467 & 0.02930 & 0.04398 & 0.05879 & 0.07371  \\
\hline
\end{tabular}
\end{center}
\end{table}

\section{Through-thickness integration of 3-D micropolar equations}
In Cartesian coordinates, the 3-D micropolar stress equilibrium equations without body forces and moments for stresses $\sigma_{ij}$ and couple-stresses $m_{ij}$ can be written as \citep{lakes1987}
\begin{equation}
\begin{aligned}
\sigma_{xx,x}+\sigma_{yx,y}+\sigma_{zx,z}&=0, \\
\sigma_{xy,x}+\sigma_{yy,y}+\sigma_{zy,z}&=0, \\
\sigma_{xz,x}+\sigma_{yz,y}+\sigma_{zz,z}&=0, \\
m_{xx,x}+m_{yx,y}+m_{zx,z}+\sigma_{yz}-\sigma_{zy}&=0, \\
m_{xy,x}+m_{yy,y}+m_{zy,z}+\sigma_{zx}-\sigma_{xz}&=0, \\
m_{xz,x}+m_{yz,y}+m_{zz,z}+\sigma_{xy}-\sigma_{yx}&=0.
\end{aligned}
\end{equation}
Analogous integration procedures with respect to the plate thickness coordinate $z$ as in the context of classical elasticity \citep{vinson2006,karttunen2017} give the 2-D micropolar plate equations
\begin{equation}
\begin{aligned}
N_{xx,x}+N_{yx,y}+\left[\sigma_{zx}\right]^{h/2}_{-h/2}&=0, \\
N_{xy,x}+N_{yy,y}+\left[\sigma_{zy}\right]^{h/2}_{-h/2}&=0, \\
M_{xx,x}+M_{yx,y}-Q_{zx}+[z\sigma_{zx}]^{h/2}_{-h/2}&=0, \\
M_{yy,y}+M_{xy,x}-Q_{zy}+[z\sigma_{zy}]^{h/2}_{-h/2}&=0, \\
Q_{xz,x}+Q_{yz,y}+[\sigma_{zz}]^{h/2}_{-h/2}&=0, \\
P_{xx,x}+P_{yx,y}+Q_{yz}-Q_{zy}+[m_{zx}]^{h/2}_{-h/2}&=0, \\
P_{xy,x}+P_{yy,y}+Q_{zx}-Q_{xz}+[m_{zy}]^{h/2}_{-h/2}&=0, \\
P_{xz,x}+P_{yz,y}+N_{xy}-N_{yx}+[m_{zz}]^{h/2}_{-h/2}&=0, \\
\end{aligned}
\end{equation}
where $h$ is the plate thickness and the last equation is related to the drilling degree of freedom that was not taken into account in the micropolar plate theory presented in Sections 2 and 3. The stress resultants are obtained from
\begin{equation}
\begin{aligned}
\left(N_{xx},N_{yy},N_{xy},N_{yx}\right)&=\int_{-h/2}^{h/2}\left(\sigma_{xx},\sigma_{yy},\sigma_{xy},\sigma_{yx}\right)dz, \\
\left(M_{xx},M_{yy},M_{xy},M_{yx}\right)&=\int_{-h/2}^{h/2}\left(z\sigma_{xx},z\sigma_{yy},z\sigma_{xy},z\sigma_{yx}\right)dz, \\
\left(Q_{xz},Q_{zx},Q_{yz},Q_{zy}\right)&=\int_{-h/2}^{h/2}\left(\sigma_{xz},\sigma_{zx},\sigma_{yz},\sigma_{zy}\right)dz, \\
\left(P_{xx},P_{yy},P_{xy},P_{yx}\right)&=\int_{-h/2}^{h/2}\left(m_{xx},m_{yy},m_{xy},m_{yx}\right)dz.
\end{aligned}
\end{equation}
\bibliographystyle{elsarticle-harv}
\bibliography{microlitera}




\end{document}